\documentclass[journal]{IEEEtran}

\usepackage{graphicx}
\usepackage{amssymb}
\usepackage{cite}
\usepackage{amsmath}
\usepackage{subcaption}
\usepackage{lipsum}
\usepackage{multicol}
\usepackage[normalem]{ulem}
\useunder{\uline}{\ul}{}
\usepackage{comment}
\usepackage{longtable}
\newcommand\ytl[2]{
\parbox[b]{8em}{\hfill{\color{cyan}\bfseries\sffamily #1}~$\cdots\cdots$~}\makebox[0pt][c]{$\bullet$}\vrule\quad \parbox[c]{4.5cm}{\vspace{7pt}\color{red!40!black!80}\raggedright\sffamily #2.\\[7pt]}\\[-3pt]}
\usepackage{supertabular}
\usepackage{algorithmicx}
\usepackage[table,xcdraw]{xcolor}
\usepackage{multirow}
\usepackage{breqn}
\usepackage{array}
\newcolumntype{C}[1]{>{\centering\let\newline\\\arraybackslash\hspace{0pt}}m{#1}}
\usepackage{comment}

\ifCLASSINFOpdf
  
\else
  
\fi

\usepackage{amsmath}
\usepackage{mathtools}

\begin{document}

\author{Arlene~John,
        Barry~Cardiff
        and~Deepu~John% <-this % stops a space
\thanks{Arlene John is with the Faculty of Electrical Engineering, Mathematics and Computer Science, University of Twente, Enschede, 7522 NB, The Netherlands. E-mail:a.john@utwente.nl}% <-this % stops a space
\thanks{Barry Cardiff and Deepu John  are with the School
of Electrical and Electronic Engineering, University College Dublin, Ireland. E-mail:\{barry.cardiff,deepu.john\}@ucd.ie}% <-this % stops a space
}

\title{A Review on Multisensor Data Fusion for Wearable Health Monitoring}

\maketitle

\begin{abstract}
The growing demand for accurate, continuous, and non-invasive health monitoring has propelled multi-sensor data fusion to the forefront of healthcare technology. This review aims to provide an overview of the development of fusion frameworks in the literature and common terminology used in fusion literature. The review introduces the fusion classification standards and methods that are most relevant from an algorithm development perspective. Applications of the reviewed fusion frameworks in fields such as defense, autonomous driving, robotics, and image fusion are also discussed to provide contextual information on the various fusion methodologies that have been developed in this field. This review provides a comprehensive analysis of multi-sensor data fusion methods applied to health monitoring systems, focusing on key algorithms, applications, challenges, and future directions. We examine commonly used fusion techniques—including Kalman filters, Bayesian networks, and machine learning models. By integrating data from various sources, these fusion approaches enhance the reliability, accuracy, and resilience of health monitoring systems. However, challenges such as data quality and differences in acquisition systems exist, calling for intelligent fusion algorithms in recent years. The review finally converges on applications of fusion algorithms in biomedical inference tasks like heartbeat detection, respiration rate estimation, sleep apnea detection, arrhythmia detection, and atrial fibrillation detection.

\end{abstract}


% \fi
\IEEEpeerreviewmaketitle

\section{Introduction}
 Healthcare expenditure has been drastically increasing worldwide over the past several decades due to an ageing population and chronic health conditions. Continuous and proactive monitoring of vital health signs outside a lab environment using wearable sensors is widely seen as the most attractive option for healthcare management. Non-invasive ambulatory physiological signal monitoring research focuses on monitoring primary physiological signals like heart rate, respiration rate, peripheral capillary oxygen saturation, blood pressure, etc and derived physiological parameters like drowsiness levels, gait, etc. However, there are several challenges involved in making wearable sensors a reality. One of the major challenges is the low quality of the signals acquired due to motion artifacts, lack of robustness to a node failure, power demands and architecture area. Corrupted signals can cause significant fluctuations in measured or derived parameters, resulting in false alarms in ambulatory health monitoring devices. This can lead to alarm fatigue and delay responses to critical events \cite{Chambrin}. In other cases, a single lead may fail, resulting in a loss of information. Additionally, due to the compact design of sensors, electrode contacts are often placed much closer together than standard measurement locations. This proximity leads to low signal amplitude, further contributing to signal quality issues and causing poor feature extraction and data interpretation \cite{Gieraltowski}.
\par Data fusion has emerged as a solution that provides improved accuracy and more precise inferences compared to those derived from a single sensor source. Multi-sensor data fusion is beneficial in scenarios when one of the sensor sources utilized in a monitoring or inference task is faulty or corrupted by noise, such that the other sensors can contribute more to the inference process. Data fusion algorithms enhance the performance of a particular task by combining information from multiple sensor sources \cite{Bruser}. 
Although data fusion is a widely used solution, no definitive methods or methodologies have been established for fusing physiological signals to enhance the quality of health monitoring or inference. This review aims to condense the diverse field of data fusion to identify relevant fusion methodologies applicable to monitoring and inference, specifically for 1-dimensional time-series data obtained from health monitoring sensors, such as electrocardiograms (ECG), photoplethsymogram (PPG), accelerometer data etc. However, with this review, the expectation is that the fusion methodologies identified could also be extended to broader applications in healthcare.

\section{Methodology}
The theory of data fusion is also highly relevant to fields such as autonomous vehicles, environmental monitoring, structural health monitoring of civil infrastructure, robotics, etc. In this review, we delve into the details of existing fusion frameworks in disparate fields to conclude a suitable framework classification strategy for the multisensor fusion works carried out in the field of biomedical signal processing. This review was carried out using a systematic search across multiple databases, including Scopus, Web of Science, PubMed, IEEEXplore, and Google Scholar. The keywords used included ``data fusion'', ``multisensor (multi-sensor) data fusion'', ``fusion methodology'', and ``fusion frameworks.'' A snowballing search on the included articles were carried out to ensure no works were missed. The included studies are discussed in this review and are used to develop or compile a preference list of fusion frameworks that are useful in classifying multisensor fusion algorithms developed for health monitoring for biomedical 1-dimensional time series signals. After the inclusion of the articles on fusion frameworks, a similar search was carried out for the monitoring applications included in this review, such as multi-sensor data fusion for heartbeat detection, heart rate estimation, respiration rate estimation, sleep apnea detection, arrhythmia detection, and atrial fibrillation detection. The inclusion criteria is that the works should include the fusion of multiple sensors for the development of a multi-sensor fusion algorithm for these applications. Fusion of different features obtained from a single sensor source is excluded.
\section{Data fusion frameworks in literature}
\label{lit_rev_theory}
This section provides a brief description of data fusion architectures discussed in the literature. Fusion methods are often categorized based on the application domain, fusion objective, sensor type, sensor suite configuration, etc. A number of data fusion frameworks have been developed over the years to aid in the identification of the most appropriate fusion methodologies for varying applications. Table \ref{timeline_table} depicts the development timeline of various fusion architectures and data fusion reviews. In this section, we elaborate on these fusion methodologies and present findings chronologically, helping the reader trace potential relationships and influences within the literature.

\begin{figure}[h]
  \centering
  \includegraphics[width=0.5\textwidth,keepaspectratio]{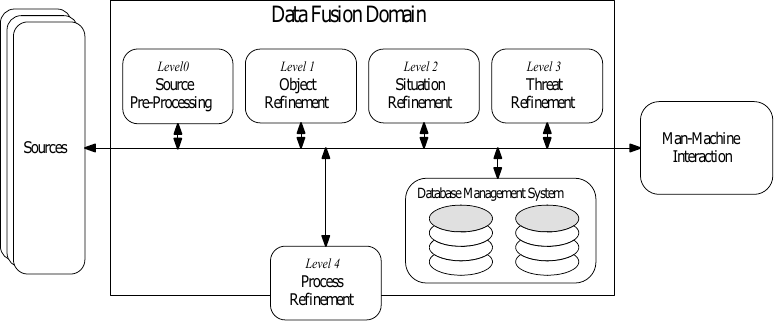}
  \caption{The JDL fusion framework. Credits: \cite{Elmenreich}.}
  \label{jdl_fig}
\end{figure}

The Joint Directors of Laboratories (JDL) data fusion framework, the oldest and most widely used framework, was developed to support advancements in military applications \cite{JDL_1987}. JDL introduces processes, functions, and different techniques with four key processes: object refinement, situation refinement, threat refinement, and process refinement (Fig. \ref{jdl_fig}). While this model has been applied in large-scale military projects, it appears to be highly specific to this domain.

\begin{table}
\caption{Timeline of the development of fusion frameworks}
\centering
\begin{minipage}[t]{.7\linewidth}
\color{gray}
\rule{\linewidth}{1pt}
\ytl{1987}{JDL \cite{JDL_1987}}
\ytl{1988}{Pau \cite{Pau_1988}. \\ Schoess and Castore \cite{Schoess_1988}. \\ Durrant-Whyte \cite{Hugh}}
\ytl{1989}{Luo and Kay \cite{Luo_1989}.\\ Thomopoulos \cite{Thomopoulos_1989}}
\ytl{1993}{Clement \textit{et al.} \cite{Clement_1993}}
\ytl{1997}{Dasarathy \cite{Dasarathy}}
\ytl{1998}{Harris \textit{et al.} \cite{Harris_1998}.\\ Pohl and Genderen \cite{Pohl_1998}}
\ytl{2000}{Bedworth and O'Brien \cite{Bedworth_2000}}
\ytl{2001}{Keithley \cite{Keithley_2001}}
\ytl{2003}{Laskey and Mahoney \cite{Laskey_2003}.\\ Chang \textit{et al.} \cite{Chang_2003}.\\ Carvalho \textit{et al.} \cite{Carvalho_2003} }
\ytl{2004}{Brooks \textit{et al.} \cite{brooks_2004}.\\ Ahmed and Pottie \cite{Ahmed_2004}.\\ Blasch \textit{et al.} \cite{Blasch_2004}}
\ytl{2005}{Esteban \textit{et al.} \cite{Esteban_2005}}
\ytl{2008}{Cohen and Edan \cite{Cohen_2008}}
\ytl{2009}{Bowman and Steinberg \textit{et al.} \cite{Bowman_2009}.\\ Llinas \cite{Llinas_2009}.\\ Dong \textit{et al.} \cite{Dong_2009}}
\ytl{2010}{Chen \cite{Chen_2010}}
\ytl{2013}{Castanedo \cite{Castanedo}}
\ytl{2017}{Fung \cite{Fung_2017}}

\bigskip
\rule{\linewidth}{1pt}%
\end{minipage}%
\label{timeline_table}
\end{table}

A data fusion framework based on a behavioral knowledge framework was proposed by Pau in \cite{Pau_1988}. The initial stages of the framework involve feature vector extraction and vector alignment to obtain a fused feature vector. A set of behavioral rules that formalize the world model is applied to the fused representation of the feature vectors. In \cite{Schoess_1988}, multiple smart sensors are serviced by a knowledge-based sensor supervisor, that uses the confidence level assigned to each sensor to process data from the sensors as an integrated group, and these multiple sensor groups are then combined to create a reconfigurable and fault-tolerant sensor fusion architecture. A data fusion classification framework was proposed by Durrant-Whyte, which was based on the relationship between data sources or sensors: complementary, redundant, and cooperative \cite{Hugh}. 

Luo and Kay proposed a fusion classification based on the information abstraction level at which fusion happens: signal-level, pixel-level, characteristic/feature-level, symbol/decision-level \cite{Luo_1989}. A comparable fusion classification approach was introduced in \cite{Thomopoulos_1989}, based on the presence of mathematical or statistical models. This method includes: (1) Signal-level fusion, where data correlation is achieved through learning in the absence of a mathematical model describing the measured phenomenon; (2) Evidence-level fusion, which combines data at various inference levels using a statistical model tailored to the user’s requirements, such as decision-making or hypothesis testing; and (3) Dynamics-level fusion, where data fusion is performed utilizing an existing mathematical model. 

Clement \textit{et al.} proposed a multi-specialist sensor fusion architecture wherein possibility theory is used to model knowledge of sensors and the uncertainty and imprecision of models in the context of image fusion in remote sensing \cite{Clement_1993}. Three kinds of specialists are present in the architecture: generic specialists (scene and conflict), semantic object specialists, and low-level specialists, and a blackboard structure with centralized control is used for fusion.

Dasarathy \cite{Dasarathy} proposed a fusion classification architecture based on the input and output information abstraction level: data in-data out, data in-feature out, feature in-feature out, feature in-decision out, decision in-decision out. Dasarathy also discusses the concept of self-improving multi-sensor fusion system architectures, wherein the central (fusion system) and local (individual sensor subsystems) decision-makers mutually enhance the other’s performance by providing reinforced learning. Fusion system architectures for environments where local decision-makers can only make limited decisions, covering only a subset of possible choices, were also discussed in\cite{Dasarathy}. 

Harris \textit{et al.} introduced a hierarchical framework known as the waterfall model \cite{Harris_1998}. In this architecture, data flows from the data level to the decision-making level, with the sensor system continuously updated via feedback from the decision-making module. The waterfall model comprises three levels of representation: (1) Level 1: Raw data is transformed to extract essential information about the environment, leveraging models of the sensors and the phenomena being measured; (2) Level 2: Features are extracted and fused, producing a list of estimates accompanied by probabilities and associated beliefs; (3) Level 3: This level connects objects to events. Pohl and Genderen \cite{Pohl_1998} reviewed data fusion methods for pixel-level image fusion and classified the various fusion algorithms based on the fusion methodology into (1) band selection methods, (2) color-related techniques, (3) statistical/ numerical methods, and (4) other combined approaches. %Growe proposed a framework based on the representation of prior knowledge through semantic nets for fusion in remote sensing images \cite{Growe_1999}.

Bedworth and O’Brien \cite{Bedworth_2000} proposed the Omnibus model, which was an approach towards creating a generalized fusion framework. This process model is a hybrid of Dasarathy's framework of I/O-based categorization \cite{Dasarathy} and the Waterfall model \cite{Harris_1998}, along with a feedback loop based on decisions and actions, inspired by military strategist John Boyd's work \textit{A discourse on winning and losing} \cite{Boyd_1987}. 

Keithley introduced a fusion evaluation methodology centered on information needs, which inherently addresses questions of `quality' \cite{Keithley_2001}. This approach employs a canonical structure that integrates quality constraints of the information needs with corresponding actions, forming a `knowledge matrix.' This matrix enables the mapping of varied information needs onto a unified framework for evaluation. The overall evaluation method is inherently multi-intelligence, with fusion serving as just one component of the broader process.

Laskey and Mahoney explore the application of probabilistic network-based knowledge and data fusion within the domain of medical diagnosis \cite{Laskey_2003}. A method to characterize the relationship between sensed information inputs and the quality of fused information output was proposed by Chang \textit{et al.} to evaluate and understand the fusion system performance \cite{Chang_2003}. A data fusion architecture utilizing the Unified Modeling Language (UML) was developed, incorporating a taxonomy based on the definitions of raw data, variables, or tasks that can be applied to network-based dynamic distributed systems \cite{Carvalho_2003}. The key advantage of this architecture lies in its adaptability, as it can be reconfigured to align with the characteristics of the measured environment and the availability of sensing units or data sources. This flexibility facilitates the creation of a generalized data fusion algorithm.

A brief overview of soft computing methodologies employed in decision-making for challenging optimization problems, particularly in the context of information fusion, was presented in \cite{brooks_2004}. These methodologies include linear programming, fuzzy logic, and artificial neural networks. The approaches discussed are considered a class of heuristics aimed at minimizing a specified objective function to derive a solution. Ahmed and Pottie proposed a probabilistic, information-processing approach to data fusion in multi-sensor networks, for which a Bayesian approach was provided and justified using information measures \cite{Ahmed_2004}. Blasch \textit{et al.} discusses the development of metrics to analyze fusion performance in dynamic situation analysis \cite{Blasch_2004}, both for user assessment and algorithm comparison. The proposed metrics for the evaluation of fusion algorithms included accuracy (error), timeliness (delay), cost (dollars), confidence (probability), and throughput (amount). 

A system-based approach for fusion architecture definition was proposed in \cite{Esteban_2005}, featuring a framework structured around three key steps in system analysis: identification, estimation, and validation.

Cohen and Edan propose a sensor fusion framework that employs a selection algorithm to identify the most reliable logical sensors in real time and determine the most appropriate algorithm for fusing data from these sensors on a robotic platform \cite{Cohen_2008}. The framework incorporates measures to assess sensor performance online. The fusion architecture prioritizes using the simplest sensor fusion algorithm alongside the most reliable sensors.

Bowman and Steinberg proposed a process for developing data fusion systems by offering guidelines for selecting fusion algorithms tailored to specific applications \cite{Bowman_2009}. They conceptualized data fusion systems engineering as a resource management process, where available techniques and design resources are utilized to achieve an objective criterion. Resource management was framed as a process of mapping the problem space to the solution space. Llinas discusses a four-step test and assessment methodology to identify the criteria, measures, and metrics to evaluate a multi-sensor fusion process performance \cite{Llinas_2009}. A review of multi-sensor satellite image fusion systems was provided in \cite{Dong_2009}, with the review focusing on various fusion frameworks and improvements and developments in existent fusion algorithms. 

explores the concept of characterizing the relationship between the sensed information inputs available to the fusion system and the quality of the fused information output, aiming to enhance the evaluation and understanding of the fusion system's performance \cite{Chen_2010}, similar to the discussion provided in \cite{Chang_2003}. The proposed fusion performance model was built upon Bayesian theory, combined with other simulation and analytical methods.

Castanedo reviews various fusion classification schemes and common fusion algorithms \cite{Castanedo}. These methods and algorithms are classified into three categories: (1) data association, (2) state estimation, and (3) decision fusion.

Fung reviews the most common fusion algorithms discussed in the literature, and it was identified that most fusion architectures are variations of (1) Kalman filtering, (2) Machine learning, (3) Bayesian inference techniques, (4) Sequential Monte Carlo (particle filtering), (5) Dempster Schafer theory of evidencing, and (6) artificial neural networks \cite{Fung_2017}.

From the history of data fusion studies, it can be observed that various fusion frameworks and classification strategies were developed, but fusion systems are heavily influenced by the application it is used for, and no single generalized fusion framework can capture the nature of all fusion algorithms used in scientific research. It can also be observed that over time, studies essentially built on existing fusion algorithms, leading to a few algorithms that end up being most commonly used. For example, fusion frameworks and architectures in the field of military systems and robotics place importance on architecture, control flow, feedback systems, etc.

This article further discusses fusion algorithm development in the field of biomedical signal processing for 1-dimensional time-series signals for health monitoring, and therefore, the fusion classification techniques used in this article are based on the Durrant-Whyte classification, Luo and Kay classification, and Dasarathy's models, which are discussed in detail next.

\subsection{Fusion classification based on the relationship between input data sources}
\begin{figure}[h]
  \centering
  \includegraphics[width=0.5\textwidth,keepaspectratio]{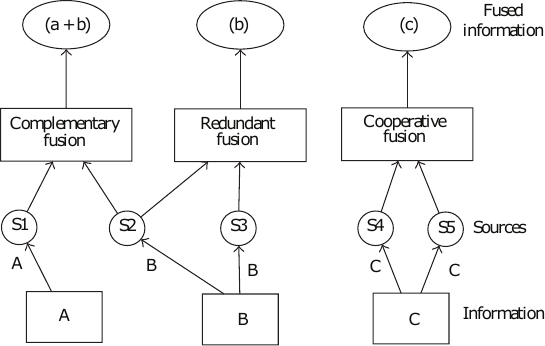}
  \caption{The Durrant-Whyte classification of fusion algorithms is based on the relationship between the data sources. Sensors $S1$ and $S2$ observe different objects, and therefore their fusion is complementary fusion. Sensors $S2$ and $S3$ observe the same object, and therefore their fusion is competitive/redundant.  Sensors $S4$ and $S5$ observe the same object, but the information provided by the two independent sensors is fused to derive information that would not be available from the single sensors. Credits: \cite{Castanedo}.}
  \label{whyte_model}
\end{figure}

A data fusion classification framework was proposed by Durrant-Whyte \cite{Hugh}, which was based on the relationship between input data sources and classified as:
\begin{enumerate}
    \item Complementary fusion: In complementary fusion, the information from the input sources offers different perspectives on the scene, allowing them to be combined to generate a more comprehensive understanding of the target. For example, ECG and blood pressure (BP) signals were fused to identify heartbeat locations in \cite{Bollepalli}.
    \item Redundant/ Competitive fusion: In redundant fusion, two or more sensors gather information about the same target, which can be combined to enhance the confidence in the target's estimation. For example, eight channels of PPG signals are fused to detect atrial fibrillation (AF) events in \cite{Shashikumar2017}.
    \item Cooperative: Cooperative fusion occurs when the provided information of the same source is integrated to create new, more complex insights than that can be obtained with a single sensor. For example, in \cite{Bonomi2016}, accelerometer data is used to remove noisy segments of ECG signals to aid in the detection of atrial fibrillation events. 
\end{enumerate}
The Durrant-Whyte classification is detailed in Fig. \ref{whyte_model}

\subsection{Fusion classification based on the fusion abstraction level}
\begin{figure}[h]
  \centering
  \includegraphics[width=0.5\textwidth,keepaspectratio]{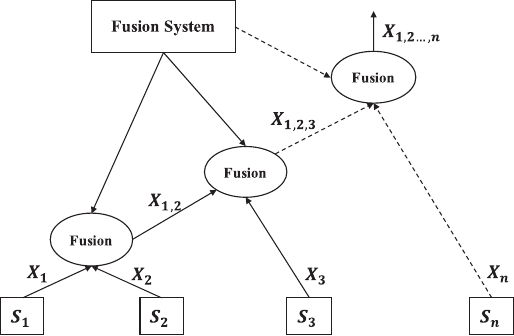}
  \caption{
Luo and Kay's classification method categorizes fusion algorithms based on the level of information abstraction at which fusion occurs, dividing them into data-level fusion, feature-level fusion, and decision-level fusion. Credits: \cite{meng_2020}.}
  \label{luo_kay_fig}
\end{figure}

Fusion algorithms were classified based on the information abstraction level at which fusion happens by Luo and Kay \cite{Luo_1989}. The classification is as shown in Fig. \ref{luo_kay_fig}.

\subsubsection{Data-level fusion}
Data-level fusion (also referred to as signal-level fusion) focuses on fusing raw signals obtained from the sensors. This method ensures that no information is lost during processing or feature extraction. However, this method can give faulty results if the signals are corrupted, which can then be corrected/adjusted later if fusion is introduced at the feature level. An example of this is image stitching for object tracking, where the input images from multiple cameras are stitched together to provide a wider field of view \cite{liu_2017}. %Signal level fusion also has the issue of spatial or temporal registration/alignment to prevent mismatch during fusion.
\subsubsection{Feature-level fusion}
Feature-level fusion involves the fusion of features extracted from the sensor sources. These features can be time-frequency images of 1-D signals such as short-term Fourier transform images or spectrogram images, edges detected in a 2D image, time-frequency information in the wavelet domain, etc. In \cite{John_heartbeat}, stationary wavelet transform (SWT) feature sequences from ECG and PPG signals were extracted to carry out a weighted sum to detect the peaks corresponding to heartbeat locations in the fused sequence to improve the accuracy of heartbeat detection. Since, the fusion process involved features extracted from the sensor input, this is an example of feature-level fusion.
\subsubsection{Decision-level fusion}
Decision-level fusion occurs at the output stage. When multiple sensor and processing units are estimating a single desired output, the reliability of the detected output can be enhanced through fusion at the decision stage, using methods such as majority voting, weighted voting, or other more advanced techniques. For example, the fusion of the R-peak (QRS peak in ECG signals) positions identified from multiple sources was used to improve the reliability of heartrate (HR) calculation in \cite{Rankawat}. Here, R-peak positions detected from ECG signals, R-peak positions detected from ECG artifacts in electromyogram (EMG), electrooculogram (EOG), and electroencephalogram (EEG) signals, and R-peak positions detected from arterial blood pressure (ABP) signals were fused using a weighted majority voting approach \cite{Rankawat}. The weights were calculated using the individual signal quality (signal quality detection methods will be discussed in Section \ref{lit_sqi}) of the various signal inputs in windows of 5-second duration.
\subsubsection{Temporal fusion}
An often overlooked aspect of the fusion process is temporal fusion, which refers to the integration of data or information collected over time. This type of fusion can take place at any of the three levels mentioned earlier and is considered orthogonal to the level-based categorization. Many other terms such as spatial and spectral fusion have been occasionally used in literature to characterize a few specific fusion methods. Temporal fusion leverages the time-based relationships between multiple sensors. An example of temporal fusion at the feature level is the use of particle filtering for extracting HR by fusing ECG, PPG, and accelerometer data \cite{Nathan}. HR was extracted from ECG using time series peak detection methods and using a time-frequency spectrogram in case of the PPG signal. Here, the most dominant frequencies close to human HRs in the accelerometer data were used to correct for errors due to motion artifacts in the proposed particle filtering-based fusion method.
\subsection{Input/Output-based characterization}
The three-level (data-feature-decision) hierarchy of fusion discussed above can be expanded into 5 fusion process I/O dependent modes, which was proposed by Dasarathy \cite{Dasarathy}:
\begin{enumerate}
    \item Data in- data out (DAI-DAO) fusion,
    \item Data in- feature out (DAI-FEO) fusion, 
    \item Feature in- Feature out (FEI-FEO) fusion, 
    \item Feature in- Decision out (FEI-DEO) fusion, 
    \item Decision in- Decision out (DEI-DEO) fusion.
\end{enumerate}
An example of each of the I/O modes of fusion is discussed below: 
\begin{figure}[h]
  \centering
  \includegraphics[width=0.5\textwidth,keepaspectratio]{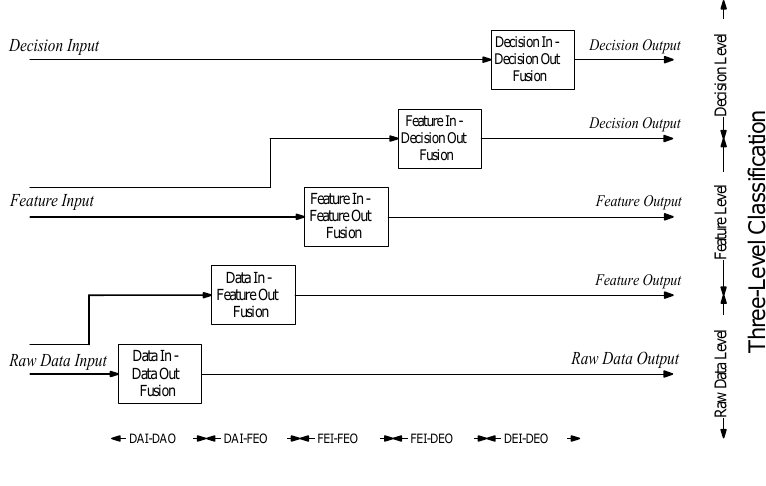}
  \caption{Dasarathy's classification of fusion based on the input-output abstraction levels. Credits: \cite{Elmenreich}.}
  \label{dasarathy_class}
\end{figure}

\subsubsection{DAI-DAO fusion}
Fusion of ECG signals with ABP and EEG signals for enhanced reliability of heartbeat detection using multimodal data association methods was proposed in \cite{Jeon}. Rule-based association models used an ABP signal to find the position of the R-peak in an ECG signal when the ECG lead was corrupted by noise. Similarly, to determine the location of the R-peak, an adaptive linear filter was used to extract the ECG signal corrupting EEG signals, when the ECG lead and ABP lead were simultaneously corrupted. The signal corruption levels were identified using signal quality indices. This is an example of DAI-DAO Fusion, with temporal fusion done based on when each signal is clean or noisy. 
\subsubsection{DAI-FEO fusion}
\par Respiratory rate detection using accelerometers and non-structured rule-based fusion of accelerometer information was discussed in \cite{Lapi}. The accelerometers placed on either side of the chest wall were used to measure the thoracic volume displacement during breathing. Here, the accelerometer data (displacements and rotations in three axes) was used as input to the fusion system, with thoracic volume displacement being the output. This is an example of DAI-FEO fusion.
%\par An attention-based time-incremental convolutional neural network that integrated CNN, recurrent cells and attention module was proposed for fusing 12-lead varied-length ECG for a 9-class arrhythmia detection problem in \cite{Yao}. The input ECG signals were directly fed into the CNN to generate a short-time pre-processed time series to capture spatial information. The generated time-series was fed into a recurrent neural network and attention mechanism to capture temporal information which would mark the signal segment where the corresponding arrhythmia detected occurred, thereby increasing interpretability of the classification process. This is an example of DAI-DEO fusion model where the raw ECG data is fed into the model to make a decision on the type of arrhythmia.
\subsubsection{FEI-FEO fusion}
\par Extraction of respiration rate intervals by fusing the signals derived from ECG and PPG signals is one of the common applications of fusion in health monitoring. Amplitude modulation, frequency modulation, and baseline wander of an ECG signal, and amplitude modulation, frequency modulation, and baseline wander of a PPG signal were fused based on a simple weighted addition in \cite{Birrenkott}. The weights for simple weighted addition were calculated using the individual signal quality of the derived signals in windows of varying durations. This is an example of FEI-FEO fusion with the modulated signals being feature inputs and the weighted sums being the outputs.
\subsubsection{FEI-DEO fusion}
\par In \cite{Kenneth}, a fuzzified rule-based method for fuzzy probability distribution of left ventricular failure, right ventricular failure, and pulmonary oedema through the fusion of multimodal cardiovascular signals, namely HR, respiration rate, peripheral oxygen saturation, and mean, systolic, and diastolic arterial blood pressures was proposed. The fuzzy probabilities of left ventricular failure, right ventricular failure, and pulmonary oedema were then used to calculate a patient deterioration rate index. This is an example of FEI-DEO fusion where the features extracted (HR, respiration rate, peripheral oxygen saturation and mean, systolic, and diastolic arterial blood pressures) from the raw sensor inputs (ECG, respiratory volume, PPG, ABP) were fed into the fuzzified rule-based model for fusion. The outputs from the fusion model were the probabilities of left ventricular failure, right ventricular failure, and pulmonary oedema, which were the requisite outputs or decisions. Similarly, a novel hierarchical heartbeat classification system was constructed to accurately classify ventricular ectopic beat (VEB) and supraventricular ectopic beat (SVEB) as discussed in \cite{Huang}. This was done by using random projections of 2 leads around the beat locations and feeding the random projections as features to an ensemble of support vector machines (SVMs)to detect VEB. Then, the ratio of the HR was compared to a predetermined threshold to detect SVEB. Here fusion was carried out by extracting features from 2 leads and feeding them to a machine learning (ML) algorithm for beat classification. This is also an example of FEI-DEO fusion.
\subsubsection{DEI-DEO fusion}
\par Bayesian fusion using the R peak to R peak (R-R) intervals (to calculate HR) obtained from 3 capacitive ECG sensors and 3 PPG sensors was proposed in \cite{Wartzek}. Instead of using a signal quality-based fusion, Bayesian fusion based on prior probabilities and likelihoods of R-R frequencies obtained per sensor is used. The difference between the estimated HR calculated using this method and the gold standard was less than ±2 bpm. This is an example of DEI-DEO Fusion as the RR-intervals were taken in as inputs for fusion to make a ‘decision’ on the most accurate R-R interval value.
\par  Dasarathy's I/O mode-based classification of fusion systems is detailed in Fig. \ref{dasarathy_class}.
\par Although the term fusion is also used in literature for the integration of multiple features extracted from the same sensor source, in this article, fusion refers to the fusion of signals, features, or decisions from multiple input sensor sources. In the next section, literature on multi-sensor fusion in the most common fusion application domains, namely robotics, military, and healthcare is reviewed. A detailed review of fusion for physiological parameter monitoring in wearable devices follows after the next section.

\section{Data fusion applications in literature}
\label{lit_app}
 Although there are a large number of publications on fusion systems for applications in various domains, they are scattered. A review of data fusion literature carried out from 2014 to 2019 focused mainly on fusion applications in robotics, military, and healthcare \cite{Jusoh_2020}. These works are discussed in detail in the next few subsections.
 \subsection{Applications in military and defense}
\par  In the military domain, major applications include 3-dimensional (3D) object detection in autonomous driving \cite{Rovid_2019}, perception and localization in autonomous driving \cite{Shahian_2019, Fayyad_2020}, moving object detection and tracking \cite{Chavez-Garcia_2016}, and obstacle detection \cite{VJohn_2019}. In \cite{Rovid_2019}, light detection and ranging (Lidar) point clouds and RGB camera images are fused through the use of deep learning networks and convolutional neural networks (CNNs) for 3D object detection in autonomous driving. In \cite{Fayyad_2020}, a review of various fusion methods for perception and localization in autonomous driving was discussed. The review focused on methods involved in the generation of local dynamic perception maps through region-based CNN (R-CNN) \cite{Girshick_2013}, spatial pyramid pooling (SPP-Net), Fast- R-CNN, Faster R-CNN, you only look once (YOLO), single-shot multibox detector (SSD), and deconvolutional single-shot detector (DSSD). The review discussed Ego-Localization and Mapping through the use of GNSS/inertial measurement unit (IMU)-based localization in which the fusion algorithms popularly used are the Kalman filter or particle filter approach, the interacting multiple model (IMM) approach, and the optimized Kalman particle swarm (OKPS) approach, recurrent neural networks (RNNs), and long short-term memory (LSTM) networks and visual-based localization algorithms using simultaneous localization and mapping (SLAM) through the fusion of stereo information or lidar data and traditional SLAM algorithms, visual odometry based localization that uses CNNs and RNNs for pose estimation. In \cite{Shahian_2019}, a fusion framework that fuses Lidar depth data and RGB camera images through an encoder-decoder-based fully convolutional neural network (FCNx) for object detection and road segmentation and an extended Kalman filter nonlinear state estimator method for moving object tracking. In \cite{Chavez-Garcia_2016}, an evidential framework (which is a generalized Bayesian framework of subjective probability) was proposed to solve the detection and tracking of moving objects problem through the fusion of radio detection and ranging (Radar), Lidar, and camera images. In \cite{VJohn_2019}, a two-stage multi-task fusion for obstacle detection is proposed, which fuses sparse radar image and camera image using CNNs, which are then used to detect big obstacles. The bottleneck features of this CNN are further fused with the raw camera image and passed through convolution layers to detect small obstacles. 
\subsection{Applications in robotics}
\par In the field of robotics, the most common applications for which fusion is used include localization \cite{Alatise_2017}, self-positioning \cite{Luo_2018}, navigation \cite{Nicosevici_2003}, and human-robot interaction \cite{Novak_2015}. Fusion of a 3-axis accelerometer and a 3-axis gyroscope with the use of an extended Kalman filter (EKF) to estimate the position and orientation of a mobile robot was proposed in \cite{Alatise_2017}. The task of self-positioning (such as walking on carpet, robot crouch, etc)/self-activity recognition is an important aspect of the robot interacting with its environment. In \cite{Luo_2018}, self-activity recognition is carried out through the use of LSTMs to model temporal information conveyed in multiple sensory streams such as the servomotor data, IMU, and pressure sensors. The features learned by the LSTMs are fused in the fully connected layer in the proposed approach. A review on underwater robot navigation was carried out in \cite{Nicosevici_2003}, where the fusion of sensors such as light amplification by stimulated emission of radiation (Laser), sound navigation and ranging (Sonar), underwater cameras, etc., are discussed and fusion algorithms can be categorized into the following types: (1) filtering and estimation-based methods, such as Kalman filtering, filtering to eliminate erroneous data, and averaging; (2) mapping-oriented techniques, including SLAM, Bayesian inference, and the extended Kalman filter for map generation; (3) behavior-based architectures, which involve two layers of sensor fusion: module-layer sensor fusion, where sensor data is combined at the signal, pixel, or feature level, and global-layer sensor fusion, where sensor information is indirectly fused by merging specific behaviors at the symbol level (e.g., obstacle avoidance, target tracking); and (4) machine learning-based systems, utilizing methods such as support vector machines and neural networks. A review on wearable sensor fusion for human-robot interaction was carried out in \cite{Novak_2015}, which discussed various rule-based fusion methods to fuse EMG, ECG, and IMUs for various applications such as gait phase recognition in lower limb robots, eye tracking, prediction of arm trajectories, and tremor suppression. 
\subsection{Applications in healthcare}
\par In healthcare, applications include monitoring and wearable technologies\cite{Nweke_2019}, wireless body sensor networks \cite{Gravina_2017}, medical diagnostics \cite{Xi_2018}, and medical imaging \cite{Bhateja_2015}. A review on wearable sensor fusion methods for human activity recognition through the fusion of ECG, PPG, HR, wearable camera, inertial sensors, etc, using ML and deep learning-based methods was discussed along with popular fusion algorithms, such as Kalman filtering and Dempster Schafer theory \cite{Nweke_2019}. In \cite{Gravina_2017}, a review on sensor fusion in body sensor networks, which is closely related to wearable sensor fusion, was carried out, which focused on body sensor networks for human activity recognition, emotion recognition, and general health monitoring. Fusion of surface EMG recordings through features extraction and classification using support vector machines to detect lower limb movement tasks was proposed in \cite{Xi_2018}. A two-stage multimodal fusion framework was proposed in \cite{Bhateja_2015} to obtain a single image for efficient clinical diagnosis. A cascaded combination of stationary wavelet transform and non-sub-sampled Contourlet transform for fusing images acquired using magnetic resonance imaging and computed tomography scan. 
\subsection{Other applications}
\par Fusion algorithms are also used in various other image fusion applications. A review on pixel-level image fusion for various applications was discussed in \cite{Bo_2010}, where image fusion algorithms were classified into simple weighted averaging fusion, multi-scale fusion which includes the pyramid scheme and wavelet fusion, point-based rules, and area-based rules with application to digital camera applications, medical imaging, surveillance and navigation, and remote sensing. In a reverse approach, tomography is introduced as a sensing task with hard-field tomography defined as a fusion process and various features of fusion algorithms were attributed to tomography images in \cite{Ozanyan_2015}.

\par Another application wherein sensor fusion plays an important role is in machine health monitoring. A review on the fusion algorithms used to fuse accelerometer data for engine fault detection in flights were carried out in \cite{Byington_2008}, which discussed various fusion algorithms using blind deconvolution, adaptive noise cancellation, active band selection/rejection, and wavelet-transform techniques to remove unwanted signal components and decompose damage event-related signals along with a discussion on approaches for normalization, resampling, and phase compensation techniques needed to align data within the same reference frame. A deep neural network approach based on stacked autoencoders is constructed to enhance the feature learning and deep feature fusion to fuse data from accelerometers, acoustic emission sensors, etc, for fault diagnosis in rotating machinery in \cite{Liu_2018}. 

\par Another interesting fusion application is in plant water stress assessment through the fusion of thermal images, aerial images, environmental data, soil and moisture data, satellite data, meterological data, and plant image data using various neural networks and ML-based methods as reviewed in \cite{Kamarudin_2021}. Multimodal fusion of thermal images and gas sensor array sequence data for gas detection using LSTM and CNNs was proposed in \cite{Narkhede_2021}. 

\par The articles reviewed in this section capture the scattered nature of data fusion literature, and it can be observed that the frameworks developed for fusion are heavily dependent on the application. However, some common trends can be observed, and few repeatedly used methods are:
\begin{enumerate}
    \item Kalman filtering for state estimation once sensor fusion problems are converted into estimation problems.
    \item Multiscale fusion of multimodal data in image fusion.
    \item Simple weighted averaging or sample-wise/pixel-wise operations in unimodal data fusion, and
    \item ML-based methods for feature fusion which also includes deep learning and CNNs with raw data inputs.
\end{enumerate}
The next section examines various fusion methodologies employed in human physiology monitoring. It evaluates whether the previously established classification effectively categorizes fusion algorithms for physiological monitoring tasks and explores whether common trends in fusion algorithms emerge across different applications in this field.

\section{Fusion approaches for physiological monitoring}
\label{lit_phy}
\begin{figure}[h]
  \centering
  \includegraphics[width=0.5\textwidth,keepaspectratio]{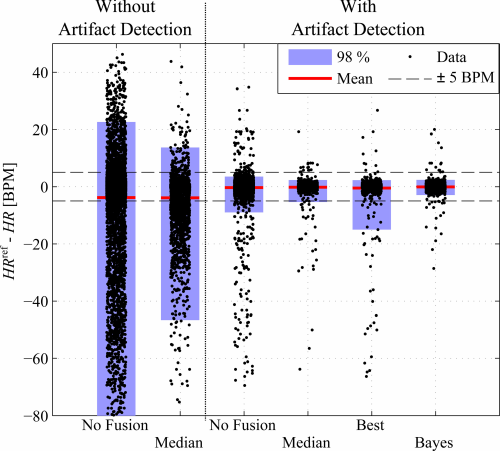}
  \caption{Gains due to fusion over single sensor methods. Credits: \cite{Wartzek}}
  \label{fus_ben}
\end{figure}

Traditional physiological signal monitoring with bedside monitors is ideal for accurate real-time inferences. The advantage of this method is that the stationary patient and sterile environment provides signals of high diagnostic quality. On the other hand, wearable devices which are more suitable for long-term ambulatory health monitoring and enhance patient mobility are highly prone to ambulatory noises of unpredictable nature. Moreover, in physiological signal monitoring, underlying physiological characteristics can be determined or inferred from a wide variety of signals. For example, heartbeats can be detected from multiple ECG signals placed on the body, from PPG sensors, ballistocardiogram (BCG) sensors, and more. Even in the context of non-wearable monitoring, video-based methods for HR estimation have been discussed in literature \cite{Balakrishnan_2013}. Therefore, multiple sensors can track the physiological characteristic of interest, and can be used in wearable devices to obtain more accurate inferences through data fusion. Moreover, fusing information from various sensors can prevent cases of inaccurate inferences due to ambulatory noise. Hence, multi-sensor data fusion is the most coherent course of action for physiological parameter tracking or estimation in wearable devices. Fig. \ref{fus_ben}, which is taken from \cite{Wartzek}, clearly illustrates the advantages of fusion compared to single-sensor (non-fusion) methods. In \cite{Wartzek}, various multi-sensor fusion algorithms were explored such as the median of the sensor readings, best sensor reading based on a previously calculated quality index, and a Bayesian approach to fusion for HR estimation. Fig. \ref{fus_ben} shows the errors in HR estimation, and it can be observed that the fusion algorithms exhibit lower errors over no fusion, especially when paired with artifact detection. Fig. \ref{fus_ben} also shows that artifact detection or fusion of signals free of artifacts is important to achieve maximum gains from fusion. The next section deals with the premise of signal quality indices (SQIs) in the context of fusing signals free of artifacts to prevent catastrophic fusion.

\subsection{The premise of Signal Quality Indicators}
\label{lit_sqi}
Fusion of the data obtained from various sensors is employed to provide a meaningful representation of the physiological data of the subjects. The acquired data is used to assess the user's fitness or health condition, and in certain cases, clinicians utilize this data for more detailed clinical analysis. The data obtained from wearable devices while subjects move through their day-to-day life is often corrupted by noise. Moreover, the placement locations of the sensors on the subjects' bodies may not be ideal for measurement. 
\par From literature, it is often observed that as the number of sensors increases, the performance of the fusion system improves. This conclusion, however, is flawed, as there are instances when the performance of a data fusion system, $F$, is actually inferior to that of the individual sensors. This phenomenon is known as catastrophic fusion and clearly should be avoided at all times \cite{mitchell}. In ambulatory health monitoring, the cause of catastrophic fusion can be attributed to spurious signals or data obtained from the sensors, as model generation and testing are often carried out in ideal or sterile conditions. Therefore, the quality of the acquired signal plays a crucial role in the accuracy of the signal analysis, particularly for wearable physiological monitoring devices that offer real-time feedback to the user. This is where sensor fusion becomes necessary to enhance inferences, with the automatic estimation of noisy or clean segments (signal quality) in continuously monitored data being crucial for health and fitness monitoring devices. This is the role of SQIs in fusion for ambulatory monitoring. 
\par The signal quality indicators represent the level of sensor uncertainty in the sensor models and are useful in preventing the phenomenon of catastrophic fusion. The use of signal quality indicators is similar to online sensor performance quantification as discussed by Cohen and Edan in \cite{Cohen_2008}. It can be observed from data fusion literature that the usage of SQIs was formalized and clearly defined for fusion algorithms in health monitoring, with various SQI algorithms developed for this purpose \cite{sqi,john_ojcas, john_csqi, john_icecs}. In the next few subsections, a review of data fusion literature for specific applications in health monitoring, such as heartbeat detection, heartrate estimation, respiration rate estimation, sleep apnea detection, arrhythmia detection, and atrial fibrillation detection is carried out. For this review, specific keywords connected to the application in conjunction with the keywords discussed in the methods section were used to carry out a systematic search across multiple databases.

\subsection{Data fusion for heartbeat detection}
\begin{figure}[h]
  \centering
  \includegraphics[width=0.5\textwidth,keepaspectratio]{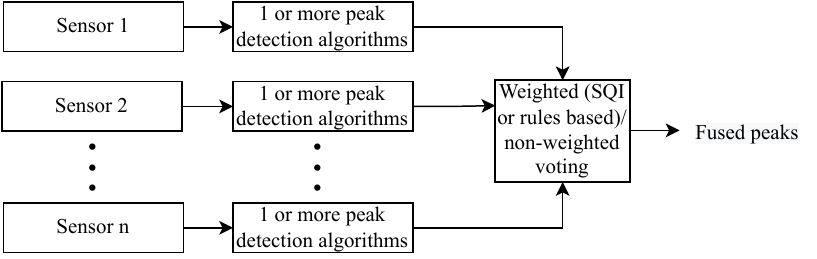}
  \caption{Summarized flow diagram of voting-based decision fusion algorithms for heartbeat detection as discussed in \cite{Zhao2018, Galeotti, Rankawat}.}
  \label{voting_beat}
\end{figure}

In the literature, most methods for heartbeat detection focus on identifying the QRS peak in ECG signals to calculate HR. This method is preferred because the QRS peak is typically easy to detect, as it accounts for 90\% of the signal energy in a beat cycle. Although most heartbeat detection algorithms focus on using ECG signals for inference, fusion with other indicative signals such as PPG, EEG, etc., are also used. The PhysioNet/CinC Challenge 2014 aimed to improve heartbeat detection using multimodal data. The challenge dataset included signals like ECG, stroke volume (SV), BP, EMG, etc., \cite{Silvai}. An example of signal-level fusion for heartbeat detection is presented in \cite{Diao}, where a novel fusion algorithm utilizing 12-lead ECG signals is introduced, based on the concept of a local weighted linear prediction algorithm. Fusion is performed to obtain a more accurate estimate of the ECG signal from 12 noisy channels taken from the CinC 2011 challenge database, which is subsequently used for HR estimation. Few works in the literature focus on voting-based methods wherein the heartbeat location is selected based on the agreement between the peaks obtained from the different sensors. In these works, the beat locations are obtained using existing peak detection algorithms like Pan-Tompkins \cite{Pan}, SQRS \cite{Engelse}, WQRS \cite{Zong}, variations of these algorithms, etc. This method is used in \cite{Zhao2018} where 12 leads of ECG signal are used to finalize the beat location, and in \cite{Galeotti}, where ECG, PPG, BP, SV signal, etc., are used for beat location and weighted voting. For the voting stage, the ECG, BP, and PPG signals were assigned higher weights due to their performance on the PhysioNet/CinC Challenge 2014 training set. These works are an example of decision-level fusion using a voting approach. A flow diagram summarizing the fusion algorithms for heartbeat detection based on voting-based methods is shown in Fig. \ref{voting_beat}. 

\par
\begin{figure}[h]
  \centering
  \includegraphics[width=0.5\textwidth,keepaspectratio]{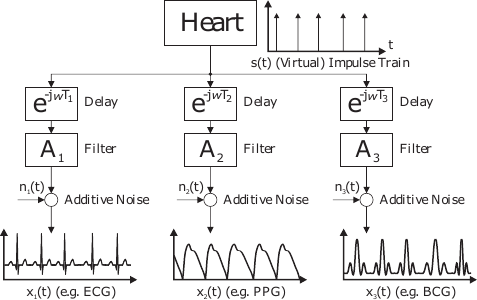}
  \caption{The blind deconvolution based source-filter approach to fusion for heartbeat detection. Credits: \cite{Antink2014}}
  \label{blind_decon}
\end{figure}

 A modeling approach where the heart is modeled as a single-input multi-output system with the outputs being ECG, PPG, and BCG signals obtained from the sensors was proposed in \cite{Antink2014}. In this approach, fusion is carried out through a blind deconvolution to estimate the heartbeat locations (Fig. \ref{blind_decon}).

\par An enhanced approach to fusion in wearable devices involves utilizing signal quality indices, thereby mitigating the risk of catastrophic fusion. Many works focus on signal quality index-based lead switching using ECG signals and other pulsatile signals like PPG, BP signal, etc. This method is used in \cite{Alistair, Vollmer2014, Mollakazemi2016} to select and switch between heartbeat locations obtained from ECG and blood pressure signals, wherein the beat locations are obtained using existing peak detection algorithms. This is an example of decision-level fusion where an SQI-based selection on the final position of heartbeats is carried out. In \cite{Rankawat}, a novel majority voting fusion algorithm based on beat SQI was proposed for robust HR estimation using both cardiovascular and non-cardiovascular signals. A modified slope sum function and the Teager-Kaiser energy operator were employed for beat detection from ECG and other non-cardiovascular signals, with signal quality indices used to switch between ECG and other pulsatile signals in \cite{Rankawat_2015}. 
\par
\begin{figure}[h]
  \centering
  \includegraphics[width=0.5\textwidth,keepaspectratio]{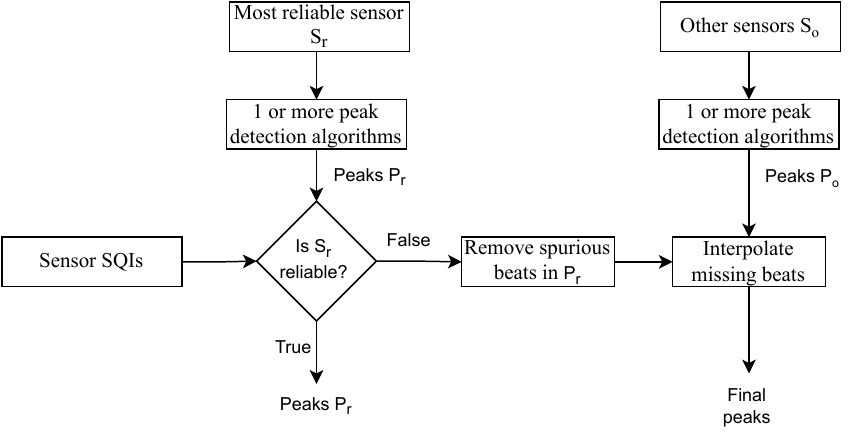}
  \caption{Summarized flow diagram of SQI/quality assessment based lead switching for heartbeat detection as discussed in \cite{Alistair, Vollmer2014, Mollakazemi2016, Rankawat_2015, Ding2014}}
  \label{sqi_lead_switching}
\end{figure}

 Some works focus on using multi-channel signals with few sensors used to detect artifacts and to exclude the signal segments that are corrupted by noise from heartbeat detection. For example, in \cite{Ding2014}, ECG signal was used for heartbeat detection and other pulsatile signals were used to check for noise. This kind of fusion can be thought of as a hybrid approach between decision fusion and feature fusion as artifacts are detected from the sensors used to detect noise, the heartbeat locations obtained from the ECG signal are eliminated. In this approach, the assumption made is that noise corrupts all the measurement channels at a time. A summarized flow diagram of the fusion algorithms for heartbeat detection based on lead-switching-based methods is shown in Fig. \ref{sqi_lead_switching}. Additionally, an SQI-based weighted fusion of the relevant SWT sequences for heartbeat detection from ECG and PPG signals was proposed in \cite{John_heartbeat}. The heartbeats were detected by carrying out peak-detection on the fused sequence. This is an example of SQI-aware fusion that does not employ lead switching.

\par
\begin{figure}[h]
  \centering
  \includegraphics[width=0.5\textwidth,keepaspectratio]{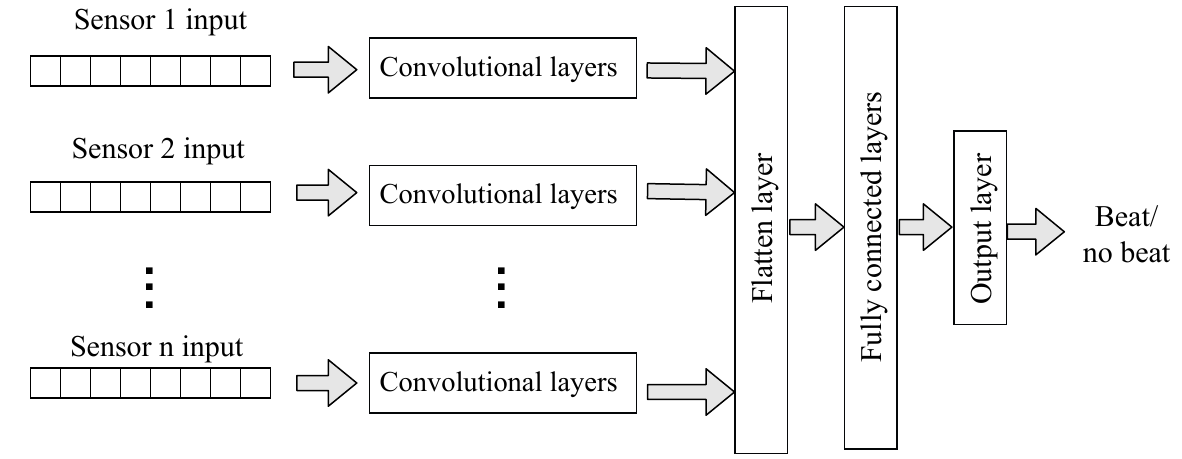}
  \caption{Summarized flow diagram of CNN-based beat detection algorithms as discussed in \cite{Bollepalli,Antink2018}. A variation of this flow is proposed in \cite{Warnecke2021}.}
  \label{cnn_beat}
\end{figure}

 With advances in ML, CNNs have become popular in processing time-series data. This has led to various works that focus on using CNNs for beat detection. This is carried out by transforming an event detection problem (heartbeat detection) into a classification problem by windowing sensor signal segments and classifying whether or not the windowed segment contained a beat approximately in the central location of the signal window. This methodology is important in fusion as any number of sensor inputs can be fed into independent convolutional layers for independent filtering of the signal segments and then fusing the transformed features in the flatten stage prior to the fully connected network for classification. This approach is used in \cite{Bollepalli} for the fusion of ECG and BP data, and in \cite{Antink2018} for the fusion of ECG and BP with application to capacitive ECG signals obtained from drivers during driving. These two works are examples of feature-level fusion algorithms as the final features after the convolutional layers are fused prior to decision-making. These works can also be considered as examples of FEI-DEO fusion. Another method of fusing image-based PPG (iPPG), PPG, BCG, and ECG signals was proposed in \cite{Warnecke2021} for the detection of heartbeats in which a hybrid approach of fusion through a voting between three different CNN networks with the networks employing early signal fusion, sensor-based late fusion (or feature fusion) with two filters for each layer for a sensor, and sensor-based late fusion (or feature fusion) with one filter for each layer for a sensor. Here, signal-level and feature-level fusion are carried out in estimating the heartbeat position in the three networks individually, and a decision fusion is carried out in the end through a voting approach. This is an example of decision fusion. A summarized flow diagram of the CNN-based fusion algorithms for heartbeat detection is shown in Fig. \ref{cnn_beat}.

A summary of the fusion algorithms discussed for heartbeat detection is provided in Table \ref{summary_table_heartbeat}. The performances of the fusion algorithms are also provided for comparison purposes. However, it should be noted that performance metrics vary from article to article. Therefore, a brief explanation of the various performance metrics used for this application is provided next:
\begin{enumerate}
\item Accuracy- the proportion of correct predictions (both true positives and true negatives) among the total number of cases examined. In the case of heart beat detection, a true positive is a correctly identified heart beat location, where an estimated location is deemed accurate if it is no further than 150 ms from the corresponding groundtruth location, as per the recommendation of the American National Standard for
ambulatory ECG analyzers (ANSI/AAMI EC38-1998).
\item Sensitivity- the proportion of correct predictions of the positives among the total number of positive cases examined. Sensitivity is also termed as recall. In the case of heart beat detection, a true positive is a correctly identified heart beat location and positives are the total number of heart beats.
\item Positive Predictive Value- the proportion of correct predictions of the positives among the total number of positives detected. Positive predictive value is also known as precision.
\item Root mean square error (RMSE)- represents the square root of the second sample moment of the differences between predicted values and observed values. In the case of heart beat detection, the quality of the prediction/ estimation is quantified based on the comparison of the estimated beat intervals to the ground truth beat intervals.
\end{enumerate}
\begin{table*}
\caption{Summary of fusion algorithms for heartbeat detection}
\label{summary_table_heartbeat}
\centering
\begin{tabular}{|p{0.04\textwidth}|p{0.18\textwidth}|p{0.25\textwidth}|p{0.2\textwidth}|p{0.2\textwidth}|}
 \hline
\textbf{Paper} & \textbf{Method} & \textbf{Fusion Architecture}                                          & \textbf{Signals}       & \textbf{Performance}                                          \\ \hline
\cite{Zhao2018}          & Voting-based decision fusion                              & Redundant, decision-level, DEI-DEO & 12 lead ECG            & Sensitivity- 93.75\%                                          \\ \hline
\cite{Galeotti}       & Voting-based decision fusion   &  Complementary, decision-level, DEI-DEO       & ECG, BP, SV, EEG, EOG  & Sensitivity- 92.80\%                                          \\ \hline
\cite{Antink2014}    & Blind deconvolution of single-input multi-output systems & complementary, feature-level, FEI-DEO  & BCG  and ECG            & RMSE-42.9 ms                                                  \\ \hline
\cite{Alistair}       & SQI-based lead switching         & complementary, decision-level, DEI-DEO\textsuperscript{1} & ECG and ABP            & Score\textsuperscript{*}- 87.88\%                                     \\ \hline
\cite{Vollmer2014}        & SQI-based lead switching                                         & complementary, decision-level, DEI-DEO\textsuperscript{1} & ECG and BP             & Sensitivity- 99.89\%                                          \\ \hline
\cite{Mollakazemi2016}    & SQI-based lead switching                                         & complementary, decision-level, DEI-DEO\textsuperscript{1}  & ECG and BP             & Sensitivity- 95.87\%                                          \\ \hline
\cite{Rankawat_2015}       & SQI-based lead switching and Taeger-Kaiser energy based R peak fusion  & complementary, decision-level, DEI-DEO\textsuperscript{1}       & EMG, EOG, and EEG      & Sensitivity- 99.81\%                                          \\ \hline
\cite{Ding2014}          & SQI-based lead switching                                         & complementary, decision-level, DEI-DEO\textsuperscript{1}  & ECG, BP, EEG, EOG, EMG & Accuracy- 81.19\%                                             \\ \hline
\cite{Rankawat}           & SQI weights-based voting for decision fusion                              & complementary, decision-level, DEI-DEO\textsuperscript{1}  & ECG, EMG, EOG, and EEG            & Score\textsuperscript{*}- 94.93\%                                          \\ \hline
\cite{John_heartbeat} & SQI-weighted fusion of feature sequence & complementary, feature-level, FEI-FEO & ECG and PPG & Sensitivity - 99.69\% \\ \hline
\cite{Bollepalli}       & CNN                                                   & complementary, signal-level, FEI-DEO  & ECG and BP             & Score\textsuperscript{*}- 94.00\%                                        \\ \hline

\cite{Antink2018}        & CNN                                          & redundant, signal-level, FEI-DEO           & 3 lead ECG             & Sensitivity- 88.00\%                                          \\ \hline
\cite{Warnecke2021}     & Hybrid CNN                                         & complementary, signal, feature, and decision-level, hybrid      & iPPG, PPG, BCG , ECG    & Average of Sensitivity and Positive Predictive Value- 94.38\% \\ 
\multicolumn{5}{|>{\scriptsize}l|}{\begin{tabular}[c]{@{}l@{}}\textsuperscript{*} Score in this table refers to the custom performance metric used in the Physionet 2014 challenge \cite{Silvai}. \\
\textsuperscript{1}It can also be noted that the SQI-based lead switching and voting methods are also an example of FEI-DEO fusion as the SQI is a feature based on which\\ the final decision is made.\end{tabular}} \\  \hline
\end{tabular}
\end{table*}
\subsection{Data fusion for heartrate estimate}
\par Continuing from heartbeat detection from the previous section, the most common health monitoring devices focus on estimating the heartrate. Fusion algorithms for heartrate estimation, where the beat locations are obtained from individual sensors and the inter-beat interval (or R peak to R peak interval (RR) in ECG signals) is calculated to generate a continuous-valued annotation along time for each sensor and then fused, are quite common in literature. Popular algorithms involve feature fusion using various algorithms to detect heartrate from a single sensor and carrying out fusion. For instance, a probabilistic model was proposed in \cite{Xie} to synthesize the HR derived from multiple QRS detection algorithms applied to ECG signals. Another example of a similar method is an unsupervised Bayesian framework that synthesizes heart rate (HR) through fusion of data from multiple heartbeat location annotators \cite{Di}. Similar approaches can be used for multisensor fusion for heartrate estimation. 
\par
\begin{figure}[h]
  \centering
  \includegraphics[width=0.5\textwidth,keepaspectratio]{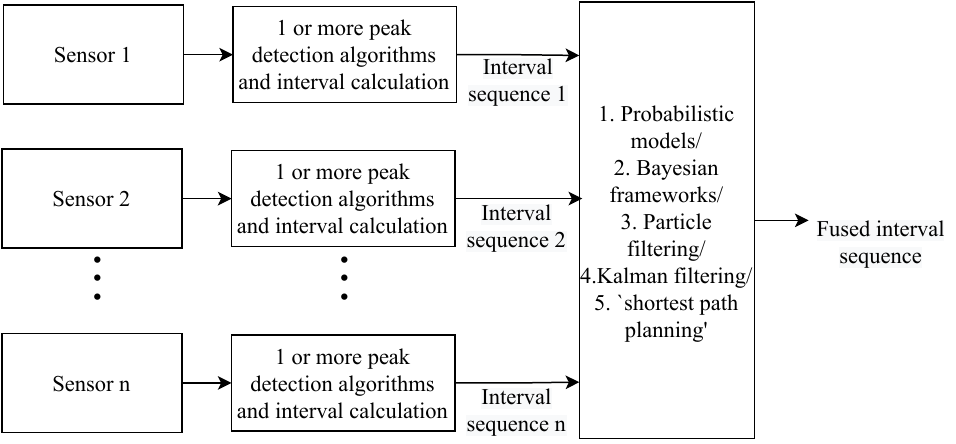}
  \caption{Summarized flow diagram of heartbeat interval fusion as discussed in \cite{Xie, Di, Nathan2, Gabriel, Wartzek, Bruser, Aygun2020, Li2015}}
  \label{rr_esti}
\end{figure}

 Other works focus on utilizing the range of HR variations from multiple sensors to help estimate the true HR. A method based on particle filtering to fuse HRs obtained from ECG and PPG signals was proposed in \cite{Nathan2}, Bayesian fusion to fuse HRs obtained from ECG, ABP, and PPG signals was proposed in \cite{Gabriel}, and to fuse HRs obtained from capacitive ECG sensors and optical pulse signals was proposed in \cite{Wartzek}. In \cite{Bruser}, two methodologies for fusing cardiac vibration signals to calculate heart rate (HR), obtained from force sensors mounted on a bed, were proposed. The first method involved analyzing the cepstrum derived from the average spectra of the individual channels, while the second method utilized Bayesian fusion with three interval estimators applied to each channel. The fusion of HR obtained from ECG and PPG signals is posed as `the shortest path detection problem' on an acyclic graph to obtain a more accurate HR \cite{Aygun2020}. A summarized flow diagram of the RR interval-based fusion algorithms for RR interval estimation is shown in Fig. \ref{rr_esti}. 

\par In \cite{Li2015}, a signal quality-aware weighted averaging of the RR intervals obtained from ECG and BP signals was proposed, where the weights are calculated based on the signal quality, Kalman filter residuals from RR estimation, and inverse quality of the other signal. This is an example of decision fusion, as the RR intervals are the final output of the fusion algorithm (Fig. \ref{rr_esti}). Another work focuses on using multi-channel signals with few sensors used to detect artifacts, and to exclude the signal segments that are corrupted by noise from RR interval calculation \cite{Huang2020}. Here, BCG signals were used for RR interval calculation and signals from other peripheral BCGs signals were used for noise estimation. An evaluation of various SQI-aware fusion methods of heartrate sequences obtained from two ECG leads was carried out in \cite{john_heartrate}. In this work, the SQIs were used as a surrogate of the input signal noise level and were used to switch between different fusion algorithms.
\par A summary of the fusion algorithms discussed for HR estimation is provided in Table \ref{summary_table_heartrate}. The performances of the fusion algorithms are also provided for comparison purposes. However, it should be noted that performance metrics vary from article to article, and a few metrics that were not discussed previously is used in the context of HR estimation. Therefore, a brief explanation of the various metrics used is provided next:
 \begin{enumerate}
 \item Mean absolute error (MAE) - represents an arithmetic average of the absolute errors between predicted values and observed values. In the case of HR estimation, the quality of the estimation is quantified based on the comparison of the estimated beat intervals to the ground truth beat intervals.
\item Relative error- represents the ratio of the absolute error of a measurement to the ground truth. 
\item Mean error- represents the arithmetic average of the errors between estimated values and observed values 
\item Correlation coefficient- is a measure of linear correlation between two sets of data. It is the ratio between the covariance of two variables and the product of their standard deviations. Here the correlation is calculated between the estimated HR and the ground truth HR.
 \end{enumerate}

\begin{table*}
\caption{Summary of fusion algorithms for heart rate estimation}
\centering
\label{summary_table_heartrate}
\begin{tabular}{|p{0.04\textwidth}|p{0.18\textwidth}|p{0.25\textwidth}|p{0.2\textwidth}|p{0.2\textwidth}|}
\hline
\textbf{Paper} & \textbf{Method} & \textbf{Fusion Architecture}                                          & \textbf{Signals}       & \textbf{Performance}                                                \\ \hline
\cite{Xie}            & Probabilistic model                  & redundant, decision-level, DEI-DEO                      & ECG                                      & RMSE- 13.96 bpm             \\ \hline
\cite{Di}             & Bayesian fusion                      & redundant, decision-level, DEI-DEO                   & ECG                                      & RMSE- 14.37 bpm             \\ \hline
\cite{Gabriel}      & Bayesian fusion                        &  complementary, decision-level, DEI-DEO                   & ECG, ABP, and PPG                        & Relative error- 23\%        \\ \hline
\cite{Wartzek}        & Bayesian fusion                       & complementary, decision-level, DEI-DEO                     & Capacitive ECG and optical pulse signals & Mean error- 0.007           \\ \hline
\cite{Bruser}         & Bayesian fusion                       & redundant, decision-level, DEI-DEO                     & BCG  sensors                    &          Relative error- 2.2\%     \\ \hline
\cite{Aygun2020}          & Direct acyclic graph based shortest path search problem    & complementary, signal-level, DAI-DAO  & ECG and PPG                              & correlation coefficient, r = 0.89 \\ \hline
\cite{Nathan2}         & Particle filtering              &                           complementary, decision-level, DEI-DEO & ECG and PPG                                      & MAE- 2 bpm              \\ \hline
\cite{Huang2020}          & Signal quality aware linear regression and Kalman  filtering & redundant, decision-level, DEI-DEO\textsuperscript{1} & BCG  sensors                              & MAE- 31 ms                  \\ \hline
\cite{john_heartrate} & Signal quality aware fusion algorithm selection & redundant, decision-level, DEI-DEO\textsuperscript{1} & ECG & MAE- 0.029 bpm \\ 
\multicolumn{5}{|>{\scriptsize}l|}{\begin{tabular}[c]{@{}l@{}}
\textsuperscript{1}It can also be noted that the SQI-based linar regression or algorithm selection are also an example of FEI-DEO fusion as the SQI is a feature based on which \\ the final decision is made.\end{tabular}} \\  \hline
\end{tabular}
\end{table*}
%\subsection{Data fusion algorithm overview}
%\subsubsection{Signal level algorithms}
%why not suitable for multimodal data
%\subsubsection{Feature level algorithms}
%suitable for multimodal data and detail on %common respresentational format
%\subsubsection{Decision level algorithms}
%details on decision level algorithms and %detail why ensembles are equivalent to %decision level algorithms

 %fusion existed for a long time-may not be biomedical fusion.
 %theoretical fusion papers
 %only temporal- no images
 %what kind of applications? arrhythmia, apnea, epilepsy? emotion detection, ECG, PPG, EEG?
 %sginal processing and ML methods
 % in ML methods- time-series images?
 % commentary on how the field is moving forward.

\subsection{Data fusion for respiratory rate detection}
Another important vital sign along with heartrate is the respiration rate (RR). Autonomic respiratory rate detection is convenient for patients with critical illness, as RR variation can signal deterioration in wards \cite{Garrido2018}, as well as the risk of patient hospitalization during remote patient monitoring \cite{Lauteslager_2024}. Respiratory modulation signals- with modulatory effects of respiration on other biomarkers- can be extracted from multiple biomedical signals, such as PPG, ECG, ABP, and the peripheral arterial tonometry (PAT) signals \cite{DF3_2021, SF2016, DF2010}. There are three main types of respiratory modulations on these signals: baseline wander (BW), amplitude modulation (AM), and frequency modulation (FM) \cite{Bookmodulations}. However, the performance of the RR estimation with the respiratory signal extracted using one modulation and a single channel signal can be easily affected by noise, leading to poor RR estimation. Due to such a limitation, data fusion techniques can be applied on multi-channel signals to improve the accuracy and robustness of RR estimation. Since various modulations can be obtained from a single sensor or multiple sensors, an important term that is commonly used in respiration rate estimation is the respiration quality index (RQI). RQIs are used to determine which of the extracted modulations are of the highest quality \cite{comb2021}. They are used in fusion literature in a manner similar to that of SQIs.

For RR estimation with a single sensor input, eg; PPG signal, the performance can be improved by fusing the RR values obtained from the three respiratory-induced modulations (BW, AM, and FM). A fusion algorithm termed as Smart Fusion employs this method, for which, if the standard deviations (SD) from the three RR values is lower than 4 breaths per minute (brpm), the mean of the three values are used as the final RR result \cite{DF5_2013}. Otherwise, the corresponding segment is labelled with a low RR estimation quality \cite{DF5_2013}. It should be noted that in \cite{DF5_2013}, Smart Fusion was employed for fusion of modulation signals obtained from a single sensor source, but the popularity of the algorithm in subsequent multi-sensor fusion literature merits mention in this review. Smart Fusion was adapted for the multi-sensor fusion scenario for RR values extracted using pulse transit time (PTT), the heart rate interval (HRI), and pulse rate interval (PRI) methods \cite{SF2016}. In \cite{DF2010}, a Kalman Filter is used to fuse the RR values extracted from ECG and PAT signals. It presents an accurate robust RR estimation framework using the Signal Purity Index (SPI) as the SQI which is based on the Hjorth descriptor to control the Kalman filter noise covariance estimate for each channel and the SQI could be adapted to fuse any number of channels' RR estimates. Besides RR values, fusion methods could also be applied on multiple modulation signals or features extracted from these signals \cite{DF3_2021}. ECG-derived respiration (EDR), PPG-derivd respiration (PDR) and BP-Derived Respiration (BDR) signals are filtered using a Kalman Filter and fused through state vector fusion here.

Another commonly used fusion method is a weighted sum for feature fusion or RR interval fusion. The weight could be calculated through multiple methods such as RQI or based on sensor reliability. In \cite{DF2017}, the weight is calculated based on the kurtosis of the power spectral density (PSD) of respiratory-related signals. A kurtosis-based weighted sum is utilized to fuse the PSD of PPG and the interval between two valleys of PPG waveforms (PPI). A similar weighted sum is applied to the two PSDs of RR-interval and the RS-amplitude signals extracted from ECG. Then, one of these two channels (PPG and ECG) is selected based on a SQI presented by Orphanidou \textit{et al.} \cite{sqi}. In \cite{Birrenkott}, three RQIs are measured based on the fast Fourier transform, autoregression, and autocorrelation methods for all the six modulations (AM, FM, BW) of ECG and PPG signals. Linear regression (LR) is used to fuse RQIs into a single RQI for each modulation. With the fused RQI, the RR estimates based on the modulation signals are merged into a single RR estimate. In \cite{DFSQI2017}, the clean heart rate variability (HRV) or Pulse Rate Variability (PRV) and AM signals extracted from noisy ECG and PPG signals are obtained based on an Ensemble Empirical Mode Decomposition (EEMD) method. Then, a quality-based fusion approach, i.e., a selection between the $RR_{E-HRV}$ and $RR_{E-AM}$ dependent on which of the two has the smallest variability in the duration of intervals between successive zero-crossings, is applied on the clean EDR signals to get the $RR_{E-FUS}$. The same fusion method is used to generate $RR_{P-FUS}$ and the final fused RR, $RR_{EP-FUS}$.

Machine learning (ML) based fusion methods are also popularly used in the domain of RR estimation. In \cite{ML2004}, a neural network approach is used to fuse three input signals, i.e., pulse wave transit time, Respiratory Sinus Arrhythmia (RSA) extracted from the ECG signal, and RSA from the PPG signal. All input points are normalized to a range of (0, 1), and the neural network is trained to predict an output of 1 when the inputs follow an inspiration. A deep learning (DL) architecture named U-Net \cite{unet} is applied to separate the demodulated respiratory (DR) signals and the non-respiratory artifacts in ECG and seismocardiogram (SCG) signals in \cite{ML2022}. Fusion is achieved by summing all DR inputs at the first convolutional layer of the U-Net architecture. Additionally, ML methods can be combined with RQI to improve the performance in RR estimation. A similar framework that combines neural networks (NNs) with a novel RQI scheme is presented in \cite{comb2021}. A bidirectional long short-term memory (BiLSTM) network is employed to predict the respiratory rate using RR and RQI features. RRs are calculated from modulation-extracted signals, and the RQI is measured based on the quantity of variation in the signal. 

A summary of the fusion algorithms discussed for respiratory estimation is provided in Table \ref{summary_table_respiratory}. The metrics used are the same as used for heartrate estimation. Another metric that was used is the error rate, which is defined as the total error out of the total measurements. Due to the difference in various fusion algorithms for respiration rate, it appeared to be quite difficult to create summary diagrams combining the various fusion algorithms utilised for this application.

\begin{table*}[]
\caption{Summary of fusion algorithms for respiratory rate estimation}
\centering
\begin{tabular}{|p{0.04\textwidth}|p{0.18\textwidth}|p{0.25\textwidth}|p{0.2\textwidth}|p{0.2\textwidth}|}
\hline
\textbf{Paper} & \textbf{Method} & \textbf{Fusion Architecture}                                          & \textbf{Signals}       & \textbf{Performance}                                                \\ \hline

\cite{SF2016}            & Smart fusion & complementary, decision-level, DEI-DEO                                        & ECG, PPG, and ABP                                      & RMSE- 1.76 brpm             \\ \hline
\cite{DF2010}        & SQI-based weights and Kalman Filter   & complementary, decision-level, DEI-DEO                                          & ECG, PPG, ABP, and PAT & RMSE- 2.72 brpm           \\ \hline

\cite{DF2017}          & SQI-based lead switching and a kurtosis-based fusion method & complementary, decision-level, DEI-DEO\textsuperscript{1} & ECG and PPG                              & -                 \\ \hline
\cite{DFSQI2017}          & SQI-based lead switching & complementary, decision-level, DEI-DEO\textsuperscript{1} & ECG and PPG                              & mean error- 1.8 brpm                  \\ \hline
\cite{ML2004}          & neural network & complementary, feature-level, FEI-DEO & ECG and PPG                              & error rate- 9\%                  \\ \hline
\cite{ML2022}          & U-Net & complementary, decision-level, DEI-DEO & ECG and SCG                              & MAE- 0.82 brpm                  \\ \hline
\cite{Birrenkott}          & RQI-based weights and linear regression & complementary, decision-level, DEI-DEO\textsuperscript{1} & ECG and PPG                              & MAE- 0.71 brpm                  \\ \hline
\cite{comb2021}          & RQI-based weights and neural network & complementary, decision-level, FEI-DEO & ECG and PPG                              & MAE- 0.638 brpm                  \\ 
\multicolumn{5}{|>{\scriptsize}l|}{\begin{tabular}[c]{@{}l@{}}
\textsuperscript{1}It can also be noted that the SQI or RQI-based lead switching or linear regression are also an example of FEI-DEO fusion as the SQI is a feature based on which \\ the final decision is made.\end{tabular}} \\  \hline
\end{tabular}
\label{summary_table_respiratory}
\end{table*}

\subsection{Data fusion for sleep apnea detection}
Sleep apnea-hypopnea syndrome is characterized by abnormal reductions or pauses in breathing during sleep and affects 10\% of middle-aged adults \cite{Peppard}. It can cause neurological arousal, impairing sleep quality and resulting in daytime sleepiness and fatigue \cite{Xie_apnea}. Monitoring sleep apnea is crucial because it helps identify disruptions in breathing that impair sleep quality, leading to effective treatment and improvement in overall health and daytime alertness. Sleep-related disorders are typically diagnosed through overnight polysomnography under clinical supervision, a process that is both expensive and often uncomfortable for patients. As a result, developing a non-intrusive, automated method for detecting sleep apnea is crucial, particularly one that can integrate seamlessly with existing smartwatch systems that monitor sleep quality. Numerous studies have demonstrated that ECG signals \cite{Atri, Billy, Urtnasan, Almazaydeh, Varon, Nguyen, Liang, Bernardini}, peripheral oxygen saturtion (SpO2) signals \cite{Mostafa2, Mostafa, Cen1}, abdominal movements \cite{Lin, Avci, Chang}, and respiratory airflow \cite{Otero, Cen1} can be used to detect sleep apnea events. The fusion of these signals for the accurate detection of sleep apnea events has also been explored in the literature.

A fusion approach using a state-space-based modeling of the cardiorespiratory system for sleep apnea detection was proposed in \cite{Gutta2018}, where the ECG and SpO2 signals were considered state inputs. In \cite{Otero}, fuzzy structural algorithms were applied to identify and characterize apnea and hypopnea episodes using respiratory airflow and SpO2 data.
\par
\begin{figure}[h]
  \centering
  \includegraphics[width=0.5\textwidth,keepaspectratio]{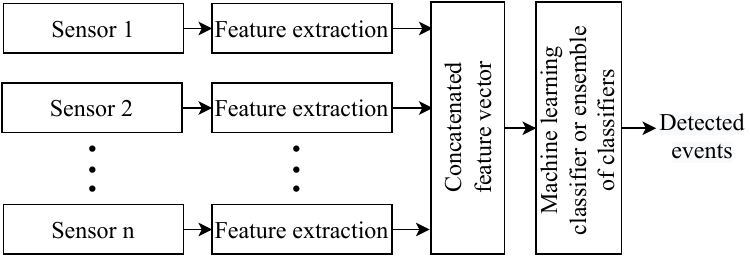}
  \caption{Summarized flow diagram of the machine learning-based fusion methods for apnea detection as discussed in \cite{Xie_apnea, Memis2017, Avci, Lin}}
  \label{apnea_ml}
\end{figure}

 ML-based models where the features from individual signals are extracted for sleep apnea detection are a popular approach. In \cite{Xie_apnea}, fusion methods combining features extracted from ECG and SpO2 signals were proposed for apnea detection, utilizing various classifiers and classifier ensembles. In \cite{Memis2017}, features derived from ECG and SpO2 signals were fed as feature inputs to a SVM for apnea detection. Wavelet domain features extracted from nasal airflow, abdominal movement, and chest movement signals (representing breathing patterns) were fused with a classifier ensemble for apnea detection in \cite{Avci}. Lin \textit{et al.} proposed an adaptive nonharmonic model to extract features associated with thoracic and abdominal movement signals collected from wearable piezoelectric bands, and used these features to classify sleep apnea events with SVMs \cite{Lin}. These studies are examples of feature-level and FEI-DEO fusion. A summarized flow diagram of the ML-based feature fusion for apnea detection is shown in Fig. \ref{apnea_ml}.

\par
\begin{figure}[h]
  \centering
  \includegraphics[width=0.5\textwidth,keepaspectratio]{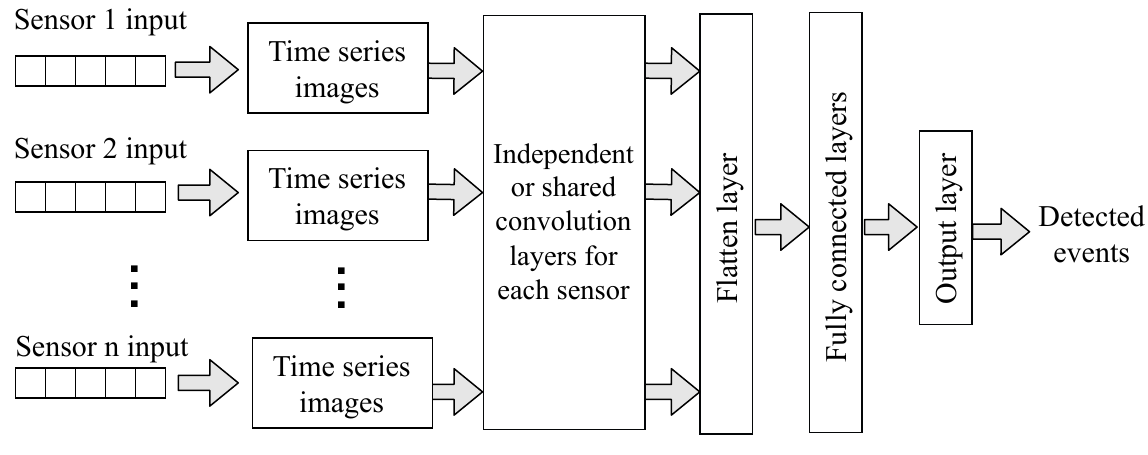}
  \caption{Summarized flow diagram of 2D-CNN based fusion methods for apnea detection (used in \cite{Cen1}).}
  \label{2d_cnn}
\end{figure}

\begin{figure}[h]
  \centering
  \includegraphics[width=0.5\textwidth,keepaspectratio]{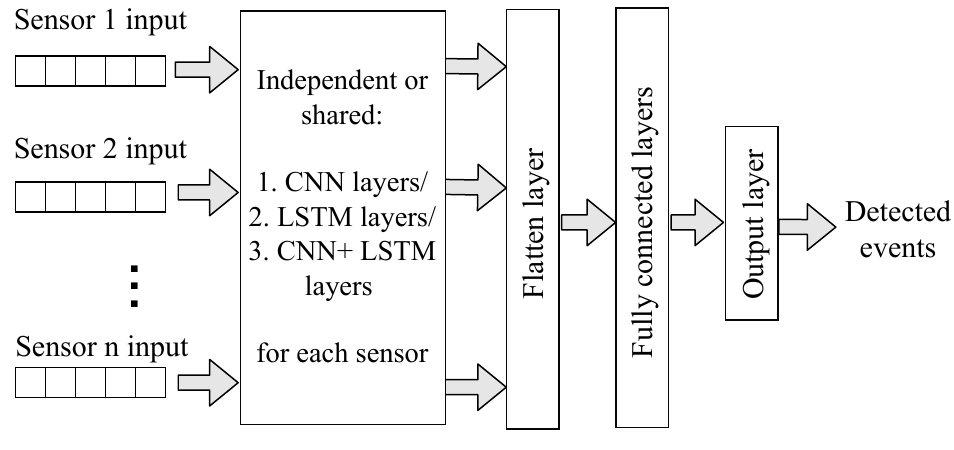}
  \caption{Summarized flow diagram of CNN/LSTM-based fusion methods for apnea detection as discussed in \cite{Bernardini, Chang, Haidar2018, Steenkiste2020, Lv2020}.}
  \label{cnn_lstm}
\end{figure}

 CNNs are very popular in the research of automatic sleep apnea detection. In literature, 1D time-series data is often transformed into time-series images for apnea detection. In \cite{Cen1}, a 2D-CNN-based method was proposed for sleep apnea detection using SpO2, oronasal airflow, ribcage, and abdominal movement data, where the signal inputs were stacked to form an image for 2D convolution (Fig. \ref{2d_cnn} shows a flow diagram of this fusion method). A method for fusing ECG and SpO2 signals using a combination of CNNs and LSTMs was proposed in \cite{Bernardini}. In \cite{Chang}, a measurement module integrating abdominal and thoracic triaxial accelerometers, ECG and SpO2 sensors was developed, and an LSTM recurrent neural network model was proposed to classify four types of sleep breathing patterns. Nasal airflow, abdominal, and thoracic channels (NAT) were used as inputs to a CNN for sleep apnea detection in \cite{Haidar2018}. A CNN-based algorithm for the fusion of ECG, SpO2, and abdominal movement signals was proposed in \cite{john_sleep_apnea}. The novelty associated with this method is the selective dropout in the CNN layers to account for differences in the sampling rate of the sensors. A novel CNN-based algorithm for sleep apnea detection through the fusion of HR signals, arterial oxygen saturation (SaO2), abdominal movement, and thoracic movement data was proposed in \cite{Steenkiste2020}. In the proposed model, the initial CNN layers are independent for the individual signals, and fusion occurs in the later stages (flatten stage), and backward shortcut connections were proposed to improve the learning of the initial stages of the model. In \cite{Lv2020}, a feature fusion network model (CNN) was proposed, where each signal record (airflow respiratory signal, thoracic respiratory signal, abdominal respiratory signal) has a multi-level feature fusion network to combine features extracted in the earlier stages and later stages. The final features of each signal record are concatenated to generate a super feature vector which is finally used for apnea detection. These studies are examples of feature-level and FEI-DEO fusion, with the variation depending only on how early in the feature extraction process fusion happens. A summarized flow diagram of the CNN or LSTM-based feature fusion methods for apnea detection is shown in Fig. \ref{cnn_lstm}.

Table \ref{summary_table_apnea} provides a summary of the fusion algorithms discussed for sleep apnea detection. It should be noted that performance metrics vary from article to article. The performance metrics include accuracy and sensitivity, where a true positive stands for a correctly identified hypopnea/apnea event. Another metric used is the area under the precision-recall curve, which shows the tradeoff between precision and recall for different thresholds. A high area under the curve represents both high recall and high precision. Precision-Recall is a useful measure of success of prediction when the classes are very imbalanced, as is the case with the sleep apnea detection problem.
\begin{table*}
\caption{Summary of fusion algorithms for sleep apnea detection}
\label{summary_table_apnea}
\centering
\begin{tabular}{|p{0.04\textwidth}|p{0.18\textwidth}|p{0.25\textwidth}|p{0.2\textwidth}|p{0.2\textwidth}|}
 \hline
\textbf{Paper} & \textbf{Method} & \textbf{Fusion Architecture}                                          & \textbf{Signals}       & \textbf{Performance}                                                                   \\ \hline
\cite{Gutta2018}          & State-space model   & complementary, feature-level, FEI-DEO         & ECG and SpO2                                                                   & Accuracy- 82.33\%                       \\ \hline
\cite{Otero}          & Fuzzy structural algorithm & complementary, feature-level, FEI-DEO  & respiratory airflow and SpO2                                                   & Sensitivity- 95\%                       \\ \hline
\cite{Xie_apnea}            & ML classifiers and ensembles & complementary, feature-level, FEI-DEO  & ECG and SpO2                                                                   & Sensitivity- 86.81\%                    \\ \hline
\cite{Memis2017}          & SVM                  &  complementary, feature-level, FEI-DEO      & ECG and SpO2                                                                   & Accuracy- 96.64\%                       \\ \hline
\cite{Avci}           & ML classifiers and ensembles & complementary, feature-level, FEI-DEO  & nasal airflow, abdominal movement, and chest movement signals                 & Accuracy- 98.68\%                       \\ \hline
\cite{Lin}            & SVM                  & complementary, feature-level, FEI-DEO       & thoracic movement and abdominal movement signals                              & Accuracy- 81.8\% $\pm$ 9.4\%                 \\ \hline
\cite{Cen1}            & 2D-CNN            &  complementary, signal-level, FEI-DEO         & SpO2 , oronasal airflow, ribcage, and abdominal movement signals               & Accuracy- 79.61\%                       \\ \hline
\cite{Bernardini}     & CNN+LSTM          &  complementary, signal-level, FEI-DEO                 & ECG and SpO2                                                                   & Accuracy- 81.50\%                       \\ \hline
\cite{Chang}          & LSTM               &  complementary, signal-level, FEI-DEO                & thoracic movement signals, SpO2 , and ECG                                      & Accuracy- 92.3\%                        \\ \hline
\cite{Haidar2018}         & CNN           &  complementary, signal-level, FEI-DEO                     & nasal airflow, abdominal movement, and thoracic movement signals              & Accuracy- 83.5\%                        \\ \hline
\cite{john_sleep_apnea} & CNN & complementary, signal-level, FEI-DEO & ECG, SpO2, and abdominal movement & Accuracy - 99.72\%
\\ \hline
\cite{Steenkiste2020}     & CNN         & complementary, signal-level, FEI-DEO                         & SaO2, HR, abdominal movement, and thoracic movement signals           & Area under precision-recall curve- 0.74 \\ \hline
\cite{Lv2020}             & CNN   & complementary, signal-level, FEI-DEO                               & airflow respiratory signal, thoracic movement, and abdominal movement signals & Accuracy- 85.5\%                        \\ \hline
\end{tabular}
\end{table*}

\subsection{Data fusion for arrhythmia detection}
Arrhythmia is a condition or problem associated with irregularities in the heartbeat, including irregular rhythms. ECGs are well-suited for arrhythmia detection because they provide detailed, real-time measurements of the heart's electrical activity\cite{Yildirim2018}. Additionally, the results of the Physionet CinC 2015 challenge showed that data fusion-based approaches are suited for arrhythmia detection, especially to reduce false alarms \cite{Clifford2015}. Various studies focusing on arrhythmia/abnormality detection through the fusion of ECG and other pulsatile signals are reviewed in this section.
\par
\begin{figure}[h]
  \centering
  \includegraphics[width=0.5\textwidth,keepaspectratio]{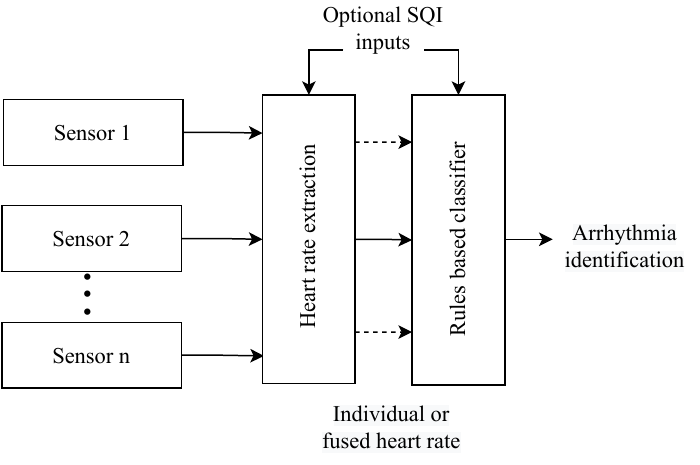}
  \caption{Summarized flow diagram of the rule-based fusion algorithms as discussed in \cite{He2015, Sadr2015, Su2018, Fallet2015, Couto2015, Teo2015, Liu2020}. Similar algorithm flow, but without SQIs, is used in \cite{Liu2015, Plesinger2015, Zong2015, Gieratowski2015}.}
  \label{arrhythmia_rules}
\end{figure}

 Many fusion algorithms focused on developing rule-based approaches to detect arrhythmia events in conjunction with the use of SQIs to reduce the cases of false alarms. In \cite{He2015}, the Kalman filter is used to extract the HR from ECG and PPG signals based on the individual SQIs. Further to which the extracted HRs and individual SQIs were used in a rule-based fusion algorithm to classify five different arrhythmias and normal sinus rhythm. In \cite{Sadr2015}, a Hilbert transform-based QRS detector was used to detect R waves from ECG signals to calculate the HR. Other open-source beat detection algorithms were used to calculate the HR from ABP and PPG signals, and the combined RR intervals obtained in conjunction with the individual signal SQIs were used to identify the type of arrhythmia. A method to trigger the arrhythmia alarm based on the agreement between the individual HRs obtained from ECG and SpO2 signals and their corresponding SQIs based on thresholding rules was proposed in \cite{Su2018}. Similar rule-based approaches for arrhythmia detection in conjunction with SQIs using ECG, PPG, and ABP signals were proposed in \cite{Fallet2015, Couto2015, Teo2015, Liu2020}. Various rule-based fusion algorithms fusing ECG and other pulsatile signals for accurate arrhythmia detection were proposed in \cite{Liu2015, Plesinger2015, Zong2015, Gieratowski2015}. A summarized flow diagram of the rule-based fusion algorithms for arrhythmia detection is shown in Fig. \ref{arrhythmia_rules}.

\par
\begin{figure}[h]
  \centering
  \includegraphics[width=0.5\textwidth,keepaspectratio]{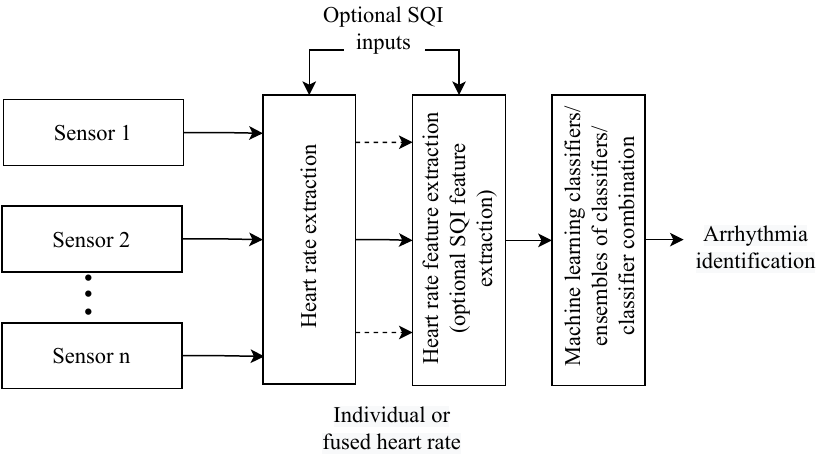}
  \caption{Summarized flow diagram of the machine learning-based fusion algorithms as discussed in \cite{Zhang2016,Caballero2015,DeChazal2004,Eerikainen2015, Srivastava2016, HoogAntink2015, Kalidas2015}.}
  \label{arrhythmia_ml}
\end{figure}

 Various multi-sensor methods for arrhythmia detection rely on machine learning models. An SQI-based peak detection algorithm is proposed to detect peaks from ECG, ABP, and PPG signals to calculate the HR, from which features are extracted to feed into an SVM classifier for arrhythmia detection in \cite{Zhang2016}. A decision-tree (DT) classifier was developed for arrhythmia classification using the HR features extracted from ECG, PPG, and ABP signals and the SQIs in \cite{Caballero2015}. Arrhythmia detection through a classifier combination of two ECG signal channels that use a linear discriminant classifier for arrhythmia detection individually is proposed in \cite{DeChazal2004}. Five different random forest classifiers were used to classify five different types of arrhythmias, where the features to the classifiers are derived from ECG, ABP, and PPG signals along with the corresponding SQIs in \cite{Eerikainen2015}, and a similar approach was discussed in \cite{Srivastava2016}. A signal similarity-based approach for interval detection from ECG and ABP/PPG signals was used to extract HR-related features for arrhythmia detection as inputs to a binary classification tree, a discriminant analysis classifier, and an SVM classifier to classify arrhythmia events and normal sinus rhythm in \cite{HoogAntink2015}. ECG and PPG signal features were used to classify five different types of arrhythmias, with each arrhythmia condition being classified with an individual SVM classifier in \cite{Kalidas2015}. A summarized flow diagram of the ML-based fusion algorithms for arrhythmia detection is shown in Fig. \ref{arrhythmia_ml}.

\par HR features and signal quality indices from ECG, PPG, and ABP signals were used as inputs to develop a case-based reasoning model for arrhythmia detection in \cite{Xu2015}. ECG and ABP signals are used as features for a k-NN classifier model for arrhythmia classification in \cite{Manna2018}. In \cite{Daluwatte2015}, various SVM-based noise classifiers for ECG, BP, and PPG signals were used to find signal segments of good reliability, and asystole, bradycardia, and tachycardia were detected from the HRs obtained from the clean signal segments. Additionally, ventricular flutter fibrillation was detected through spectrum analysis of the ECG signal, and a heartbeat classifier based on QRS template matching was used to detect ventricular tachycardia.

\par
\begin{figure}[h]
  \centering
  \includegraphics[width=0.5\textwidth,keepaspectratio]{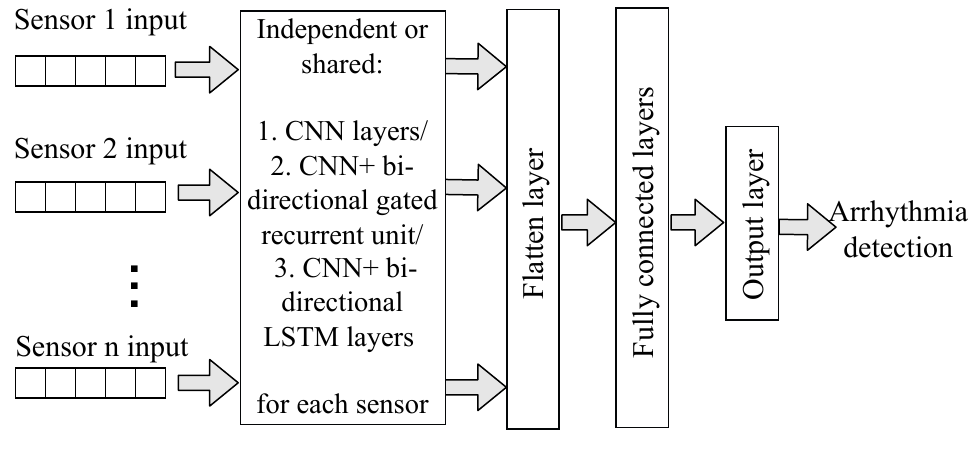}
  \caption{Summarized flow diagram of the deep-learning based fusion algorithms as discussed in \cite{Li2019,Zhang2021, Wang2019, Ye2020}.}
  \label{cnn_lstm_gru_arr}
\end{figure}

 CNN-based models are also commonly studied for arrhythmia detection and analysis. In \cite{Li2019}, a CNN-based fusion algorithm using 2-lead ECG signals, where the initial layers of the CNN for both leads do not interact and get fused only in the final few layers was proposed for distinguishing between 5 different arrhythmia types. A novel network called MLBF-Net that uses CNNs in conjunction with bidirectional gated recurrent units to extract feature maps from 12-lead ECG signals for fusion for arrhythmia detection of five different arrhythmia types was proposed in \cite{Zhang2021}. A similar architecture is proposed in \cite{Wang2019} for the fusion of 12-lead ECG signals with skip connection operation to fuse different levels of features extracted at different stages of the CNN, and channel-wise attention modules are used for effectively extracting the features learned at the different stages to classify five different arrhythmia types. Another algorithm that extracts local features using CNNs and then extracts temporal features by bi-directional LSTM for arrhythmia classification for 12 leads separately was proposed in \cite{Ye2020}, where an eXtreme Gradient Boosting (XGBoost) classifier is used to fuse the 12-lead models to distinguish between five different arrhythmia types. A summarized flow diagram of the deep learning-based fusion algorithms for arrhythmia detection is shown in Fig. \ref{cnn_lstm_gru_arr}.

A method for the detection of premature ventricular contraction through the mean fusion of the wavelet image of 6 ECG leads and feeding the Tucker-decomposed images of the inverse wavelet transform of the fused fusion image to a CNN was proposed in \cite{Hoang2019} (the flow of this method is similar to as shown in Fig. \ref{2d_cnn}). A summary of the fusion algorithms discussed for arrhythmia detection is provided in Table \ref{table_arrhythmia}. The performances of the fusion algorithms are also provided for comparison purposes. However, it should be noted that performance metrics vary from article to article, and a few metrics not discussed previously in this review are used in the context of arrhythmia detection. Therefore, a brief explanation of the various metrics used for this application is provided next:
 \begin{enumerate}
 \item False alarm suppression rate- the proportion of true negatives detected from amongst the total number of negatives, which is also known as specificity. True negatives correspond to normal sinus rhythm in case of arrhythmia detection. False alarm suppression is the terminology used when the work focuses on reducing the number of false arrhythmia alarms.
\item F1-score- the harmonic mean of sensitivity (recall) and positive predictive value (precision).
 \end{enumerate} 

\begin{table*}

\caption{Summary of fusion algorithms for arrhythmia detection}
\label{table_arrhythmia}
\centering
\begin{tabular}{|p{0.04\textwidth}|p{0.18\textwidth}|p{0.25\textwidth}|p{0.2\textwidth}|p{0.2\textwidth}|}

\hline
\textbf{Paper} & \textbf{Method}         & \textbf{Fusion-architecture}                                   & \textbf{Signals}  & \textbf{Performance}                              \\ \hline
\cite{He2015}             & SQI and HR based rules                            & complementary, feature-level, FEI-DEO & ECG and ABP & Sensitivity- 65\%                                 \\ \hline
\cite{Sadr2015}           & SQI and HR based rules                            & complementary, feature-level, FEI-DEO & ECG, PPG, and ABP & Score\textsuperscript{*}- 74.03                                      \\ \hline
\cite{Su2018}             & SQI and HR based rules                            & complementary, feature-level, FEI-DEO & ECG and SpO2       & False alarm suppression- 50\%                     \\ \hline
\cite{Fallet2015}         & SQI and HR based rules                            & complementary, feature-level, FEI-DEO & ECG, PPG, and ABP & Score\textsuperscript{*}- 76.11                                      \\ \hline
\cite{Couto2015}          & SQI and HR based rules                            & complementary, feature-level, FEI-DEO & ECG, PPG, and ABP & Score\textsuperscript{*}- 78.65                                      \\ \hline
\cite{Teo2015}            & SQI and HR based rules                            & complementary, feature-level, FEI-DEO & ECG, PPG, and ABP & Score\textsuperscript{*}- 69.92                                      \\ \hline
\cite{Liu2020}       & SQI and HR based rules   & complementary, feature-level, FEI-DEO                          & ECG, PPG, and ABP & False alarm suppression- 56.7\%                   \\ \hline
\cite{Liu2015}       & Rules based      & complementary, feature-level, FEI-DEO                                          & ECG, PPG, and ABP & Score\textsuperscript{*}- 75.91                                      \\ \hline
\cite{Plesinger2015}      & Rules based                                               & complementary, feature-level, FEI-DEO & ECG, PPG, and ABP & Score\textsuperscript{*}- 81.39                                      \\ \hline
\cite{Zong2015}           & Rules based                                               & complementary, feature-level, FEI-DEO & ECG, PPG, and ABP & Score\textsuperscript{*}- 70.10                                      \\ \hline
\cite{Gieratowski2015}   & Rules based                                               & complementary, feature-level, FEI-DEO & ECG, PPG, and ABP & Score\textsuperscript{*}- 57.72                                      \\ \hline
\cite{Zhang2016}          & SVM                                                       & complementary, feature-level, FEI-DEO & ECG, PPG, and ABP & Score\textsuperscript{*}- 84.4                                       \\ \hline
\cite{Caballero2015}      & DT & complementary, feature-level, FEI-DEO & ECG, PPG, and ABP & Score\textsuperscript{*}- 65.19                                      \\ \hline
\cite{DeChazal2004}      & Linear discriminant and voting                            & redundant, feature-level, FEI-DEO & 2 channel ECG     & SVEB sensitivity- 75.9\%, VEB sensitivity- 77.7\% \\ \hline
\cite{Eerikainen2015}    & ML classifiers along with SQI-based features & complementary, feature-level, FEI-DEO & ECG, PPG, and ABP & Score\textsuperscript{*}- 75.54                                      \\ \hline
\cite{Srivastava2016}     & ML classifiers along with SQI-based features & complementary, feature-level, FEI-DEO & ECG, PPG, and ABP & Accuracy- 83.96\%                                 \\ \hline
\cite{HoogAntink2015}    & DT, Discriminant analysis, and SVM              & complementary, feature-level, FEI-DEO & ECG and PPG/ABP   & Score- 75.55                                      \\ \hline
\cite{Kalidas2015}        & SVM                                                        & complementary, feature-level, FEI-DEO & ECG and PPG       & Sensitivity- 96\%                                 \\ \hline
\cite{Xu2015}             & Case- based reasoning model                                & complementary, feature-level, FEI-DEO & ECG, PPG, and ABP & Sensitivity- 83\%                                 \\ \hline
\cite{Manna2018}          & k-NN & complementary, feature-level, FEI-DEO & ECG and ABP       & Accuracy- 99.95\%                                 \\ \hline
\cite{Daluwatte2015}      & SVM                                                        & complementary, feature-level, FEI-DEO & ECG, PPG, and ABP & Score\textsuperscript{*}- 70.2                                       \\ \hline

\cite{Li2019}             & CNN                                                       & redundant, signal-level, FEI-DEO & 2-lead ECG        & SVEB sensitivity- 81.1\%, VEB sensitivity- 97.7\% \\ \hline
\cite{Zhang2021}          & MLBF-net + CNN                                            & redundant, signal-level and feature-level, DAI-FEO and FEI-DEO & 12-lead ECG       & F1-score- 0.855                                   \\ \hline
\cite{Wang2019}           & CNN with channel-wise attention                           & redundant, signal-level, FEI-DEO & 12-lead ECG       & F1-score- 0.813                                   \\ \hline
\cite{Ye2020}             & CNN+ Bi-directional LSTMs                                 & redundant, signal-level, FEI-DEO & 12-lead ECG       & F1-score- 0.812                                   \\ \hline
\cite{Hoang2019}          & CNN                                                       & redundant, signal-level, FEI-DEO & 6-lead ECG        & Sensitivity- 78.60\%                              \\ 
\multicolumn{5}{|l|}{\begin{tabular}[c]{@{}l@{}}\textsuperscript{*} Score in this table refers to the custom performance metric used in the Physionet 2015 challenge \cite{Clifford2015} \end{tabular}} \\  \hline
\end{tabular}
\end{table*}

\subsection{Data fusion for atrial fibrillation detection}
Atrial fibrillation (AF) is the most prevalent cardiac arrhythmia affecting 1\% to 2\% of the general population. It is characterized by rapid and disorganized atrial activation leading to impaired atrial function \cite{Pellman2015}. AF can be detected from single-lead short-term ECG recordings in clinical settings and in home environments \cite{Fan2018, Hermans_2021}. Literature on the use of data fusion using multiple ECG signals or a combination of ECG and other pulsatile signals for AF detection is sparse, despite the popularity of multisensor fusion in arrhythmia detection. In this section, we discuss few data fusion-based AF detection algorithms.
\par
\begin{figure}[h]
  \centering
  \includegraphics[width=0.5\textwidth,keepaspectratio]{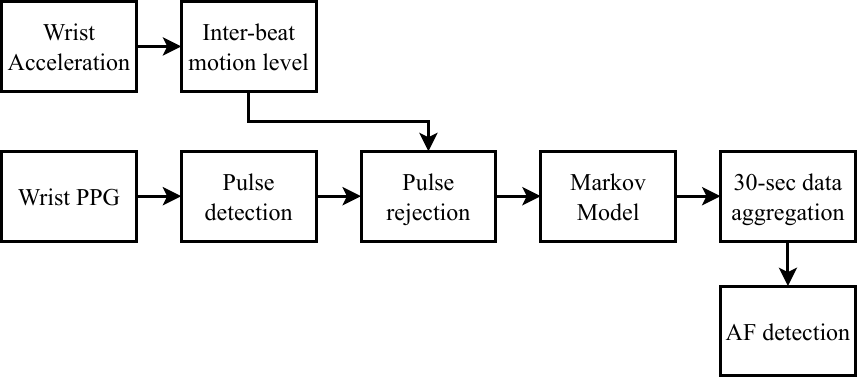}
  \caption{The Markov model-based classifier for AF detection discussed in \cite{Bonomi2016}. Adapted from \cite{Bonomi2016}.}
  \label{af_markov}
\end{figure}

 In \cite{Bonomi2016}, a Markov model-based classifier that used inter-pulse interval features from pulses detected from the ECG signal was proposed to detect AF, while using the wrist accelerometer data to reject noisy pulses  (Fig. \ref{af_markov}).

\par
\begin{figure}[h]
  \centering
  \includegraphics[width=0.5\textwidth,keepaspectratio]{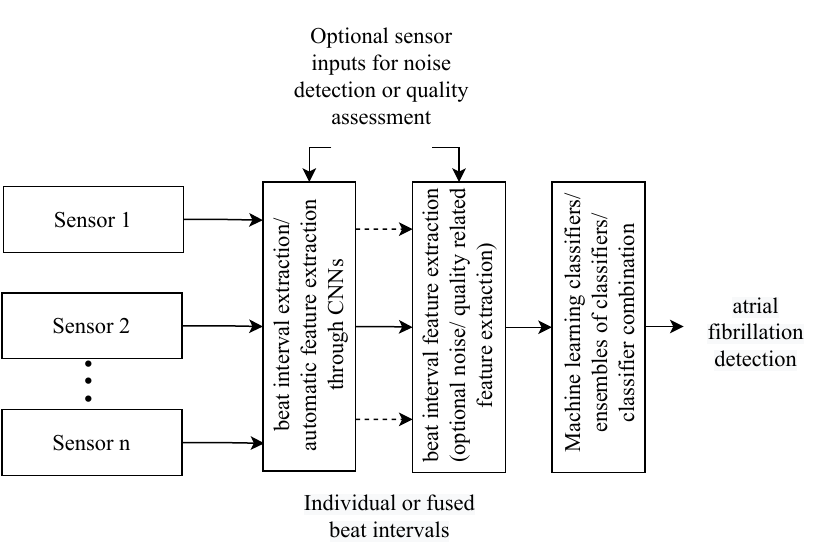}
  \caption{Summarized flow diagram of the machine learning-based fusion algorithms for AF detection as discussed in \cite{Zhu2021, Eerikainen2020, Shashikumar2017}.}
  \label{afib_ml}
\end{figure}

 A few ML-based approaches for AF detection are also discussed in the literature. A hybrid classifier that leveraged an ML classifier and a statistical rule-based classifier for AF detection was proposed in \cite{Zhu2021}, where the inter-beat intervals obtained from the PPG signal and related features were used for classification along with PPG SQIs as well as motion detection from the accelerometer data were used to eliminate noisy inter-beat intervals from the decision classifier. In \cite{Eerikainen2020}, a similar approach was used to distinguish between AF, atrial flutter, and normal sinus rhythm using other waveform features extracted from the PPG signal as well as the accelerometer data using a random forest classifier. An elastic net logic was used to detect AF using a combined feature vector obtained from a CNN network (whose input was 8-channel PPG recordings) as well as features obtained from 8-channel PPG recordings and its SQIs in \cite{Shashikumar2017}. A summarized flow diagram of the ML-based fusion algorithms for AF detection is shown in Fig. \ref{afib_ml}.

A summary of the fusion algorithms discussed for AF detection is provided in Table \ref{summary_table_afib}.

\begin{table*}[]
\centering
\caption{Summary of fusion algorithms for atrial fibrillation detection}
\begin{tabular}{|p{0.04\textwidth}|p{0.18\textwidth}|p{0.25\textwidth}|p{0.2\textwidth}|p{0.2\textwidth}|}
\hline
\textbf{Paper} & \textbf{Method} & \textbf{Fusion Architecture}                                          & \textbf{Signals}       & \textbf{Performance}                                                   \\ \hline
\cite{Bonomi2016}         & Markov model                                                           & cooperative, feature-level, FEI-FEO & PPG and accelerometer signals & Sensitivity- 97\% $\pm$ 2 \% \\ \hline
\cite{Zhu2021}            & ML classifier+ rules-based decisions &  cooperative, feature-level, FEI-DEO                      & PPG and accelerometer signals & Sensitivity- 87.8\%     \\ \hline
\cite{Eerikainen2020}     & Random forests                                                         & cooperative, feature-level, FEI-DEO  & PPG and accelerometer signals & Sensitivity- 98.2\%     \\ \hline
\cite{Shashikumar2017}    & ML classifiers using features obtained from CNNs and SQIs & redundant, feature-level, FEI-DEO  & 8- channel PPG                & Accuracy- 91.8\%        \\ \hline
\end{tabular}
\label{summary_table_afib}
\end{table*}

Now that we have thoroughly examined the key findings and insights from existing research on multisensor data fusion for specific applications, it is now essential to transition from analyzing individual study results to discussing their broader implications, challenges, and potential future directions within the field of physiological monitoring, which is discussed next. 
\section{Discussions}
\subsection{Challenges}
Although the various works discussed in this review individually address many challenges associated with multisensor data fusion for wearables, the primary challenges still remain, with most works providing solutions for specific applications. Ensuring high data quality is a primary challenge in multisensor fusion for wearables-based health monitoring, as data can often be noisy due to environmental factors or patient movement. Accurately fusing these data sources of uncertain quality requires algorithms capable of managing ambiguity and providing robust insights for healthcare applications. Uncertainty handling is critical in clinical settings, where decisions based on fused data can have significant implications for patient care. The use of SQIs have been found to tackle this issue with sufficient reliability, however, a generalised approach for all applications is not available. Additionally, health monitoring involves various data types—such as physiological signals, motion data, etc—that differ in format, precision, and sampling rates. Integrating these heterogeneous data sources requires sophisticated methods that can handle inconsistencies and align diverse data sources accurately for meaningful fusion. Some fusion works have proposed methods to address such challenges \cite{John_heartbeat, john_sleep_apnea}, but such considerations and approaches are not yet mainstream.

Real-time data fusion can be computationally intensive, especially when using advanced algorithms on wearable devices with limited processing power and battery life. Low latency and energy efficiency are crucial for continuous health monitoring applications, where data needs to be processed quickly without excessive power consumption. As these systems scale to integrate multiple data streams, managing and analyzing the high data volume adds to the computational demands, challenging the system’s ability to provide timely health insights. Some fusion works have explored the computational cost associated with increasing the number of sensors and explored opportunities to reduce computational costs \cite{john_sleep_apnea}. Additionally the use of SQIs also add to the computational overhead and the development of energy-efficient SQI algorithms are an important research direction to enable multisensor daa fusion \cite{john_icecs}.

In real-world applications, data may also be intermittently unavailable due to connectivity issues or sensor malfunctions, resulting in missing or incomplete datasets. Developing methods that can align temporally misaligned data and interpolate or manage missing values is essential to maintain the reliability of health monitoring systems.

\subsection{Future directions}
The future of multisensor data fusion for health monitoring shows significant promise for advancement. For example,  through the use of machine learning and deep learning models, such as convolutional neural networks (CNNs) and recurrent neural networks (RNNs), the handling of complex, high-dimensional data and adapting dynamically to real-time variations in sensor quality can be improved. Furthermore, developments in explainable AI (XAI) techniques are crucial to bringing multisensor data fusion in remote patient monitoring to support clinical decision-making \cite{Gerdes_2024}. Developing uncertainty modeling approaches, such as Bayesian methods, can also improve the reliability of predictions by managing ambiguous or incomplete data.

Data quality remains a fundamental priority for accurate health monitoring. Improvements in data pre-processing, such as adaptive filtering \cite{john_sg}, can enhance data quality prior to fusion. Incorporating methods to handle missing data through the development of robust imputation techniques are crucial for reliable remote patient monitoring. Additionally, due to the lack of large reliable datasets for training fusion models, especially with the restrictions placed on data privacy, federated learning, which allows model training across multiple devices without centralizing data, can also play a transformative role \cite{Wang_2023}. By processing data locally, federated learning enhances data privacy and security, enabling more personalized and adaptive health monitoring systems without compromising patient confidentiality. 

Personalized health monitoring represents a valuable future direction for multisensor remote patient monitoring systems. Adaptive models that account for individual patient characteristics can improve monitoring accuracy, as they adjust data fusion techniques based on each patient’s unique context and health history. Additionally, advances in the field of Human-in-the-loop AI can support the personalisation of these algorithms through input from the patient and their healthcare practitioner, further reducing responsibility concerns \cite{responsibility} associated with remote patient monitoring solutions.

\section{Conclusion}
\label{lit_disc}
In this article, a comprehensive review of fusion frameworks and classification criteria developed over the years was conducted. While numerous fusion frameworks and classifications have been proposed, this review focuses on those relevant to biomedical fusion algorithm development, such as the Durrant-Whyte model \cite{Hugh}, the Dasarathy model \cite{Dasarathy}, and the Luo and Kay classification \cite{Luo_1989} to assist in categorizing the algorithms reviewed. Various applications of fusion algorithms and methods in non-biomedical contexts were also discussed initially, to highlight their influence on the development of fusion algorithms for biomedical applications.
From the study of various fusion applications, several commonly used methods were identified in the literature:
\begin{enumerate}
\item Kalman filtering or its variants for state estimation, particularly when sensor fusion problems are framed as estimation problems.
\item Multiscale fusion of multimodal data in image fusion.
\item Simple weighted averaging or sample-wise/pixel-wise operations for unimodal data fusion.
\item Machine learning (ML)-based methods for feature fusion, including deep learning approaches such as convolutional neural networks (CNNs) with raw data inputs.
\end{enumerate}
The review also covered various fusion methodologies employed in physiological signal monitoring applications, including heartbeat detection, heart rate estimation, respiration rate estimation, sleep apnea detection, arrhythmia detection, and atrial fibrillation detection. Based on the application-specific analysis of fusion techniques, several commonalities in methodologies were observed. Accordingly, fusion algorithms from the reviewed literature can broadly be categorized as follows:
\begin{enumerate}
\item State estimation algorithms, utilizing Kalman filtering, Bayesian inference, particle filtering, and other probabilistic models.
\item Rule-based fusion algorithms.
\item Signal quality index-based algorithms.
\item ML or deep learning methods for feature fusion.
\item CNN-based fusion, where multi-sensor data is input into a network, and the fusion process is learned by the network itself.
\end{enumerate}
From the fusion techniques discussed, it is evident that most algorithms are application-dependent. It is also notable that the fusion architectures outlined in Section \ref{lit_rev_theory} do not explicitly account for the quality of the signals being fused. This is a critical concern for biomedical wearable devices due to the challenges posed by ambulatory noise, the miniaturized construction of sensors, and sensor placement on the body away from the primary location of interest. Consequently, SQI-based fusion algorithms are particularly significant in the context of biomedical wearable devices, as discussed in \ref{lit_phy}. Within the scope of this review, the challenges associated with multisensor data fusion in the current healthcare landscape were enumerated, and the potential avenues for future research to advance this field were highlighted.

\bibliographystyle{IEEEtran}
\bibliography{bibliography}

\end{document}